\documentclass{article}
\usepackage{arxiv}
\usepackage{cite}
\usepackage{amssymb}
\usepackage{amsfonts}
\usepackage{tikz}
\usepackage{pgfplots}
\usepackage[T1]{fontenc}
\usepackage{amsmath}
\usepackage{subfigure}
\usepackage{mathtools}
\usepackage{balance}
\usepackage{xspace}
\usepackage{xcolor}
\usepackage{algorithm}
\usepackage{algorithmicx}
\usepackage{algpseudocode}
\usepackage{algcompatible}
\usepackage{lmodern}
\usepackage{url}
\usepackage[british]{babel}
\usetikzlibrary{positioning}
\algrenewcommand\alglinenumber[1]{\scriptsize #1:}

\newcommand{\chain}{{SharPer}\xspace}
\newcommand{\Chain}{{SHARPER}\xspace}
\newcommand{\REQ}{\textsf{REQUEST}\xspace}

\newcommand{\ONE}{\textsf{PROPOSE}\xspace}
\newcommand{\one}{{\sf \small propose}\xspace}
\newcommand{\ONEF}{\textsf{SUPER-PROPOSE}\xspace}
\newcommand{\onef}{{\sf \small super-propose}\xspace}
\newcommand{\TWO}{\textsf{ACCEPT}\xspace}
\newcommand{\two}{{\sf \small accept}\xspace}
\newcommand{\TWOF}{\textsf{SUPER-ACCEPT}\xspace}
\newcommand{\twof}{{\sf \small super-accept}\xspace}
\newcommand{\TWOQ}{\textsf{ACCEPT-QUERY}\xspace}
\newcommand{\twoq}{{\sf \small accept-query}\xspace}
\newcommand{\THREE}{\textsf{COMMIT}\xspace}
\newcommand{\three}{{\sf \small commit}\xspace}
\newcommand{\FOUR}{\textsf{COMMITTED}\xspace}
\newcommand{\four}{{\sf \small committed}\xspace}
\newcommand{\UNK}{\textsf{UNKNOWN}\xspace}
\newcommand{\unk}{{\sf \small unknown}\xspace}
\newcommand{\PRE}{\textsf{PREPARE}\xspace}
\newcommand{\pre}{{\sf \small prepare}\xspace}
\newcommand{\PRO}{\textsf{PROMISE}\xspace}
\newcommand{\pro}{{\sf \small promise}\xspace}

\newcommand{\reply}{{\sf \small reply}\xspace}
\newcommand{\vchange}{{\sf \small view-change}\xspace}
\newcommand{\VCHANGE}{\textsf{VIEW-CHANGE}\xspace}
\newcommand{\checkp}{\textsf{checkpoint}\xspace}

\newcommand{\newv}{{\sf \small new-view}\xspace}
\newcommand{\NEWV}{\textsf{NEW-VIEW}\xspace}

\newtheorem{metalemma}{Lemma}[section]

\newtheorem{Lemma}[metalemma]{Lemma}

\newenvironment{lmm}{\begin{Lemma}\em}{\end{Lemma}}

\newenvironment{proof}{\noindent{\bf Proof:}\rm}

\title{\chain: Sharding Permissioned Blockchains Over Network Clusters}

\author{
Mohammad Javad Amiri \qquad Divyakant Agrawal \qquad Amr El Abbadi\\
Department of Computer Science, University of California Santa Barbara\\
Santa Barbara, California\\
\{amiri, agrawal, amr\}@cs.ucsb.edu
\vspace{2em}
}

\begin{document}
\maketitle

\begin{abstract}
Scalability is one of the main roadblocks to business adoption of blockchain systems.
Despite recent intensive research on using sharding techniques to enhance the scalability of blockchain systems,
existing solutions
do not efficiently address cross-shard transactions.
In this paper, we introduce {\em \chain}, a permissioned blockchain system that
improves scalability by 
clustering (partitioning) the nodes and assigning different data shards to different clusters
where each data shard is replicated on the nodes of a cluster.
\chain supports both intra-shard and cross-shard transactions and processes
intra-shard transactions of different clusters as well as
cross-shard transactions with non-overlapping clusters simultaneously.
In \chain, the blockchain ledger is formed as a directed acyclic graph where
each cluster maintains {\em only} a view of the ledger.
\chain also incorporates a {\em flattened} protocol to establish consensus 
among clusters on the order of cross-shard transactions.
The experimental results reveal the efficiency of \chain in terms of
performance and scalability especially in 
workloads with a low percentage of cross-shard transactions.
\end{abstract}
\section{Introduction}\label{sec:intro}

Blockchain is a distributed data structure for recording transactions
maintained by nodes without a central authority \cite{cachin2017blockchain}.
Blockchain systems are classified into two categories: {\em permissionless} systems and {\em permissioned} systems.
While in a permissionless blockchain system, e.g., Bitcoin \cite{nakamoto2008bitcoin},
the network is public, and anyone can participate without a specific identity,
a {\em permissioned} blockchain,
e.g., Hyperledger Fabric \cite{androulaki2018hyperledger},
consists of a set of known, identified nodes which might not fully trust each other.
In permissionless blockchain systems, consensus on the order of transactions is achieved through mining
whereas in permissioned blockchain systems,
asynchronous fault-tolerant protocols are used to guarantee safety.
Fault-tolerant protocols mostly rely on either $3f{+}1$ Byzantine or $2f{+}1$ crash-only nodes
to overcome the simultaneous failure of any $f$ nodes.

Scalability is one of the main obstacles to business adoption of blockchain systems.
Scalability is the ability of a blockchain system to process an increasing number of transactions
by adding resources to the system.
The scalability of blockchain systems has been addressed in several studies using different 
on-chain, e.g., increasing the block size, and
off-chain, e.g., Lightning Networks \cite{miller2019sprites}\cite{poon2016bitcoin}, techniques.
Increasing the block size, however, increases both the propagation time and the verification time of the block which
makes operating full nodes more expensive, and this in turn could cause less decentralization in the network \cite{croman2016scaling}.
Off-chain solutions also suffer from security issues \cite{seres2019topological} especially denial-of-service attacks
\footnote{{\url{https://www.trustnodes.com/2018/03/21/lightning-network-ddos-sends-20-nodes}}}.

Partitioning the data into multiple shards that are maintained by different subsets of non-malicious nodes
is a proven approach to improve the scalability of distributed databases \cite{corbett2013spanner}.
In such an approach, the performance of the database scales linearly with the number of nodes.
Recently, sharding has been utilized by several approaches in the presence of Byzantine nodes
in both permissionless and permissioned blockchain systems.
Sharded permissionless blockchains, e.g.,
Elastico \cite{luu2016secure}, OmniLedger \cite{kokoris2018omniledger}, and Rapidchain \cite{zamani2018rapidchain},
ensure {\em probabilistic} correctness by randomly assigning nodes to {\em committees} (partitions)
resulting in a uniform distribution of faulty nodes in committees.
OmniLedger and Rapidchain also support
cross-shard transactions using Byzantine consensus protocols.

Sharding techniques have also been used by different permissioned blockchains, e.g.,
Fabric \cite{androulaki2018hyperledger}, Cosmos \cite{cosmos18},
RSCoin \cite{george2015centrally}, and AHL \cite{dang2018towards}.
In Fabric, channels are introduced to shard the system.
A channel is a partitioned state of the full system that is
autonomously managed by a (logically) separate set of nodes, but is still aware
of the bigger system it belongs to \cite{androulaki2018channels}.
By using channels, Fabric is able to process intra-shard transactions efficiently.  However,
processing any cross-shard transaction needs
either the existence of a trusted channel among the participants or 
an atomic commit protocol \cite{androulaki2018channels}.
Cosmos \cite{cosmos18} introduces
Inter-Blockchain Communication (IBC)
to initiate cross-blockchain operations.
Interacting chains in IBC, however, must be aware of the state of each other which requires
establishing a bidirectional trusted channel between two blockchains.
In AHL\cite{dang2018towards}, Dang et al. employ a trusted hardware
(the technique that is presented in
\cite{chun2007attested}\cite{veronese2013efficient}\cite{veronese2010ebawa})
to decrease the number of required nodes within each committee.
AHL randomly assigns nodes to the committees and
ensures safety with a high probability if each committee consists of $80$ nodes
(instead of ${\sim} 600$ nodes in OmniLedger).
Nevertheless, running Byzantine fault-tolerant protocols
among $80$ nodes results in high latency.
In addition, in AHL \cite{dang2018towards}, consensus on the order of cross-shard transactions
not only requires an extra set of nodes (called a {\em reference committee}),
but also results in a large number of communication phases.
Furthermore, since a single reference committee processes cross-shard transactions,
AHL is not able to process cross-shard transactions with non-overlapping committees in parallel.

In many systems, especially permissioned blockchains, the number of available nodes is much larger than $3f+1$.
In such systems, using all the nodes to establish consensus degrades
performance since more messages are being exchanged without providing improved resiliency, e.g.,
in PBFT \cite{castro1999practical}, the number of message exchanges is quadratic in terms of the number of nodes.
Different techniques have been presented to address this issue.
In the active/passive replication technique,
the protocol relies only on $3f{+}1$ active nodes to establish consensus whereas
FaB \cite{martin2006fast} uses $5f{+}1$ replicas to establish consensus
in two phases instead of three as in PBFT.
Similar techniques have been presented 
for crash failures to use $3f{+}1$ replicas instead of $2f{+}1$ \cite{lamport2006fast}\cite{brasileiro2001consensus}.
However, such techniques do not utilize the extra nodes efficiently
when a very high percentage of nodes are non-faulty.

In our previous work \cite{amiri2019sharding}, we presented a model including a blockchain ledger
for sharded permissioned blockchains.
In this paper, we expand this model by first, introducing consensus protocols to order 
both intra- and cross-shard transactions
on either crash-only or Byzantine nodes and
second, designing a sharded permissioned blockchain system, {\em \chain}, to
improve scalability.
\chain partitions the nodes into {\em clusters} of either
$2f+1$ crash-only or $3f+1$ Byzantine nodes to guarantee safety and
can be used specifically in networks with very high percentage of non-faulty nodes.

\chain assigns data shards to the clusters where
each cluster processes the transactions that access its corresponding shard.
If a transaction accesses only a single shard, i.e., an {\em intra-shard transaction}, the corresponding cluster 
orders and executes the transaction locally.
As a result, intra-shard transactions of different clusters are independent of each other,
and can be processed in parallel.
However, for a {\em cross-shard transaction}, agreement among {\em all} and {\em only} involved clusters is required.
Nevertheless, if two cross-shard transactions have no overlapping clusters, they still can be processed in parallel.
Since the ordering of different transactions might be performed in parallel and the system
includes cross-shard transactions, the blockchain ledger of \chain is represented as a {\em directed acyclic graph}
including all intra- and cross-shard transactions.
Nonetheless, for the sake of performance,
the blockchain ledger is {\em not maintained} by any node and
nodes of each cluster maintain their own {\em view} of the ledger
including its intra-shard transactions and the cross-shard transactions that the cluster is involved in.
The main contributions of this paper are:

\begin{itemize}
    \item \chain, a permissioned blockchain system that supports the concurrent processing of transactions
    by clustering the nodes into clusters and sharding the data and the blockchain ledger.
    \chain supports both intra-shard and cross-shard transactions.
    
    \item Two {\em flattened} consensus protocols for ordering cross-shard
    transactions among all and only the involved clusters in networks consisting of
    either crash-only or Byzantine nodes. The protocols order
    cross-shard transactions with non-overlapping clusters in parallel.    
\end{itemize}

The rest of this paper is organized as follows.
The \chain model is introduced in Section~\ref{sec:model}.
Sections~\ref{sec:cons-crash} and~\ref{sec:cons-byz} show how consensus works in \chain.
Section~\ref{sec:eval}  presents a performance evaluation of \chain.
Section~\ref{sec:related} discusses related work, and
Section~\ref{sec:conc} concludes the paper.

\section{The \Chain Model}\label{sec:model}

\chain is a permissioned blockchain system
designed specifically to achieve high scalability in networks with a very large percentage of non-faulty nodes.
\chain partitions the nodes into clusters and assigns a data shard to each cluster.
Each node, in addition to a data shard, stores a view of the blockchain ledger.
In this section,
we first present the \chain infrastructure and show how clusters and shards are formed.
Then, the blockchain ledger is introduced.
Finally, the transaction model is discussed.

\subsection{\chain Infrastructure}

\chain consists of a set of nodes in an asynchronous distributed system.
Nodes in \chain either follow the crash or Byzantine failure model.
In the crash failure model, nodes operate at arbitrary speed,
may fail by stopping, and may restart, however they may not collude, lie, or
otherwise attempt to subvert the protocol.
Whereas, in the Byzantine failure model,
faulty nodes may exhibit arbitrary, potentially malicious, behavior.
Crash fault-tolerant protocols, e.g., Paxos \cite{lamport2001paxos},
guarantee safety in an asynchronous network using $2f{+}1$ crash-only nodes
to overcome the simultaneous crash failure of any $f$ nodes while
in Byzantine fault-tolerant protocols, e.g., PBFT \cite{castro1999practical},
$3f{+}1$ nodes are usually needed to provide the safety property in the presence of $f$ malicious nodes \cite{bracha1985asynchronous}.
In a blockchain system,
the maximum number of simultaneous failures, $f$, can be specified based on the characteristics of nodes
and historical data.

\chain uses point-to-point
bi-directional communication channels to connect nodes.
Network channels are pairwise authenticated, which guarantees
that a malicious node cannot forge a message from a correct node,
i.e., if node $i$ receives a message $m$ in the incoming link from node $j$,
then node $j$ must have sent message $m$ to $i$.
Furthermore, messages might contain public-key signatures and message digests \cite{castro1999practical}.
A {\em message digest} is a numeric representation of the contents of a message
produced by collision-resistant hash functions.
Message digests are used to protect the integrity of 
a message and detect changes and alterations to any part of the message.
We denote a message $m$ signed by replica $r$ as
$\langle m \rangle_{\sigma_r}$ and
the digest of a message $m$ by $D(m)$.
For signature verification, we assume that
all nodes have access to the public keys of all other nodes.
We assume that a strong adversary can coordinate malicious nodes and
delay communication to compromise the replicated service.
However, the adversary cannot subvert standard cryptographic assumptions
about collision-resistant hashes, encryption, and signatures.

\subsection{Cluster and Shard Formation}

\begin{figure}[t]
\begin{center}
\includegraphics[width= \linewidth]{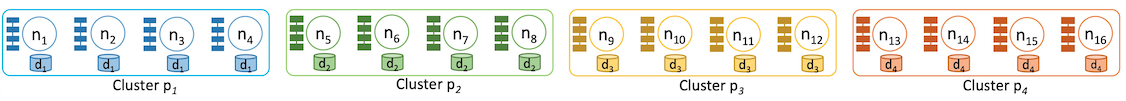}
\caption{The infrastructure of \chain with $16$ Byzantine nodes where $f=1$}
\label{fig:arch}
\end{center}
\end{figure}

In existing sharded permissionless blockchain systems,
e.g., OmniLedger \cite{kokoris2018omniledger},
nodes are assigned to clusters (committees) randomly. In such systems,
to ensure safety, i.e., each cluster includes at most one-third faulty nodes, with a high probability ($1{-}2^{-20}$),
each cluster consists of hundreds of nodes.
In the permissioned blockchain system AHL \cite{dang2018towards}, safety is ensured with the same probability
with clusters of only $80$ nodes using trusted hardware, 
however, as discussed earlier,
running fault-tolerant protocols among $80$ nodes results in high latency.
In \chain, on the other hand, the number of nodes, $N$, is assumed to be much larger than $3f+1$
(or $2f+1$ if nodes are crash-only), thus, nodes are partitioned
into clusters each large enough to tolerate $f$ failures.
As a result and in contrast to AHL, \chain provides a {\em deterministic} safety guarantee (not a probabilistic one), hence
there is no need to reconfigure clusters or assign nodes to clusters in a random manner.
Note that both the probabilistic approach and trusted hardware technique can also be utilized in \chain
resulting in enhanced performance.

In \chain and in the presence of crash-only nodes,
each cluster includes $2f+1$ nodes (the last cluster might include more nodes) and
similarly, in the Byzantine failure model, each cluster includes $3f+1$ nodes.
Nodes are assigned to the clusters mainly based on their geographical distance,
i.e., nodes that are close to each other are assigned to the same cluster
to reduce the latency of communication within a cluster.
We denote the set of clusters by $P= \{p_1, p_2, ...\}$.
If nodes are crash-only, the number of clusters, $|P|$, is equal to $\frac{N}{2f+1}$.
Similarly, in the presence of Byzantine nodes, the number of clusters, $|P|$, is $\frac{N}{3f+1}$.
The number of clusters in \chain, indeed, depends on the number of nodes, number of failures,
and the failure model of nodes. As a result, the lower the percentage of faulty nodes,
the more the number of clusters.
Since there are $|P|$ clusters, the data is also sharded into $|P|$ shards,
thus each cluster maintains a single data shard that
is replicated on the nodes of the cluster.
We denote shards by $d_1$, ..., $d_{|P|}$ where
each shard $d_i$ is replicated over the nodes of cluster $p_i$.

Figure~\ref{fig:arch} illustrates the \chain infrastructure for a blockchain system
consisting of $16$ nodes following Byzantine failure model where $f=1$.
As can be seen, the system consists of four clusters ($|P| {=} \frac{16}{4}$) of size four ($3f{+}1$).
The data is sharded into four shards where each shard $d_i$ is replicated on the nodes of cluster $p_i$.
Nodes within each cluster, in addition to a data shard, store a view of the blockchain ledger.

An appropriate sharding needs to be {\em workload-aware}, i.e., have prior knowledge of the data and
how the data is accessed by different transactions.
Workload-aware data sharding increases the probability of
maintaining the records which are accessed by a single transaction in the same shard \cite{curino2010schism}.
Different approaches have been proposed to minimize the number of distributed transactions in a sharded system \cite{pavlo2012skew},
nevertheless, there might still be a portion of transactions that accesses records from different shards.
As a result, \chain supports two types of transactions:
{\em intra-shard} transactions that access the records within a shard and
{\em cross-shard} transactions that accesses records from different shards.

\begin{figure}[t] \center
\includegraphics[width= 0.7 \linewidth]{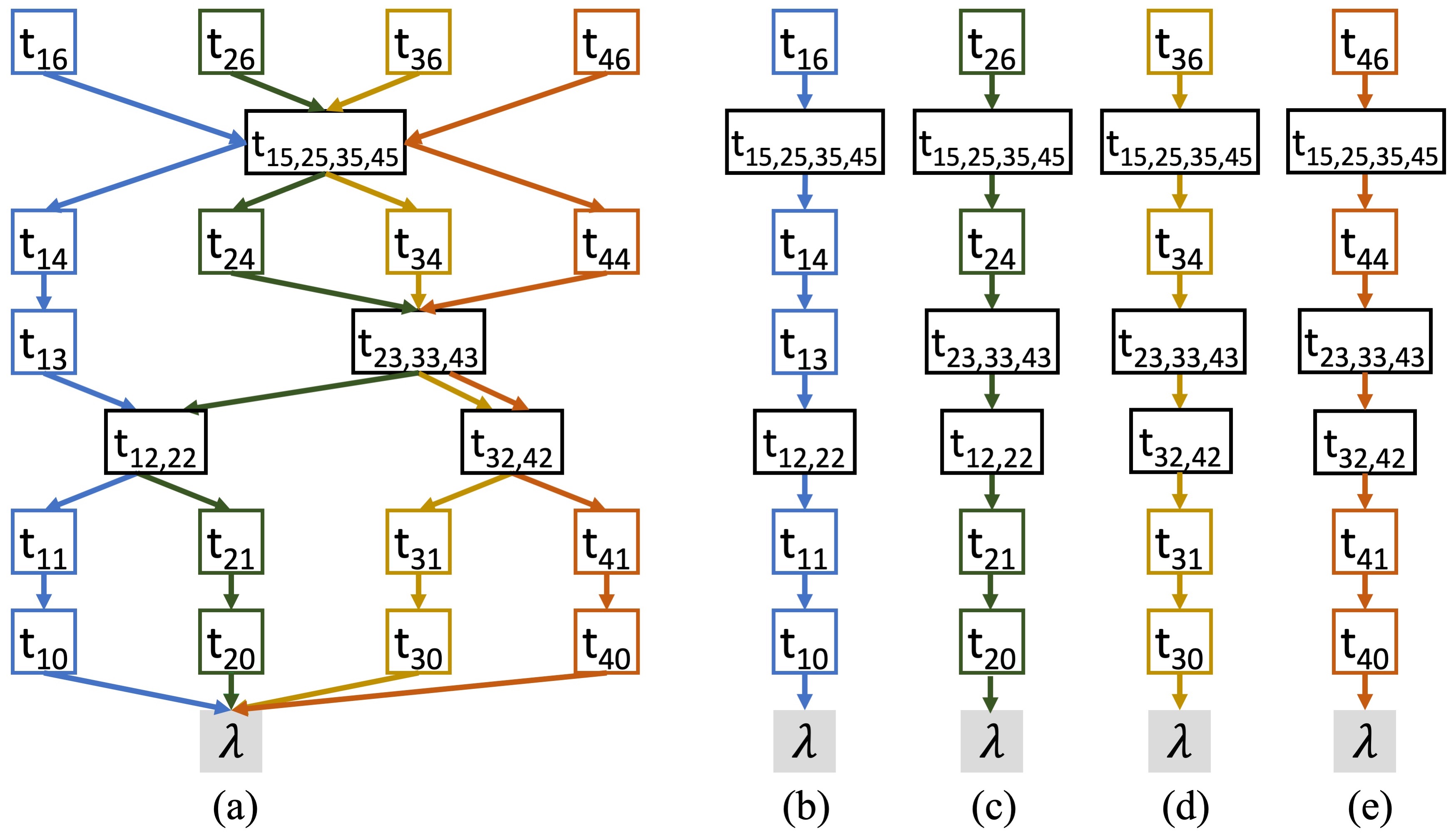}
\caption{(a): A blockchain ledger consisting of four shards,
(b), (c), (d), and (e): The views of the blockchain from the four different shards}
\label{fig:ledger}
\end{figure}

\subsection{Blockchain Ledger}

Blockchain systems record transactions in the form of a hash chain
in an append-only data structure, called the {\em blockchain ledger}.
Originally, a blockchain ledger was proposed to support cryptocurrencies as an ordered list of blocks where each block
includes a batch of transactions and the cryptographic hash of the prior block \cite{nakamoto2008bitcoin}.
While in permissionless blockchain systems, batching the transactions into blocks
amortizes the cost of cryptography, e.g., solving the cryptographic puzzles in Bitcoin, and
makes data transfers more efficient in a large geo-distributed setting,
in permissioned blockchains,
since proof-of-work is not required, as shown in \cite{istvan2018streamchain},
batching transactions into blocks decreases performance.
Thus, in \chain, each block consists of a single transaction.
In \chain, each data shard is replicated on all nodes of a cluster.
As a result, to ensure data consistency, a total order among transactions (both intra- and cross-shard)
that access the same data shard is needed.
The total order of transactions in the blockchain ledger is captured by
{\em chaining} transaction blocks (we assume each block consists of a single transaction) together, i.e.,
each block includes a sequence number or the cryptographic hash of the previous transaction block.
Since more than one cluster is involved in each cross-shard transaction,
similar to \cite{amiri2019caper} \cite{amiri2019sharding},
the ledger is formed as a {\em directed acyclic graph}.
The ledger also includes
a unique initialization block, called the {\em genesis} block.

Fig.~\ref{fig:ledger}(a) shows a blockchain ledger created in the \chain model
consisting of four clusters $p_1$, $p_2$, $p_3$, and $p_4$ (data shards $d_1$, $d_2$, $d_3$, and $d_4$).
In this figure, $\lambda$ is the genesis block of the blockchain.
Intra- and cross-shard transactions are also specified.
For example, 
$t_{10}$, $t_{11}$, $t_{13}$, $t_{14}$, and $t_{16}$
are the intra-shard transactions of cluster $p_1$.
Each cross-shard transaction is labeled with $t_{o_1,.., o_k}$ where
$k$ is the number of involved clusters and
$o_i$ indicates the order of the transaction among the transactions of the $i^{\text{th}}$ involved cluster.
This is needed to ensure that cross-shard transactions are ordered correctly
with regard to the intra-cluster transactions of all involved clusters.
For example, $t_{12,22}$ and $t_{15,25,35,45}$ are two cross-shard transactions
where $t_{12,22}$ accesses data shards $d_1$ and $d_2$, whereas
$t_{15,25,35,45}$ accesses all four data shards.

As can be seen, there is a total order among the transactions (both intra- and cross-shard) that access a data shard,
e.g., $t_{10}$, $t_{11}$, $t_{12,22}$, $t_{13}$, $t_{14}$, $t_{15,25,35,45}$, and $t_{16}$
are chained together.
In addition, intra-shard transactions of different clusters can be added to the blockchain ledger in parallel, e.g.,
$t_{11}$, $t_{21}$, $t_{31}$, and $t_{41}$ can be processed by different clusters in parallel.
Similarly, if two cross-shard transactions access disjoint subsets of shards,
they can be added to the ledger in parallel as well, e.g., $t_{12,22}$ and $t_{32,42}$.

In \chain, the entire blockchain ledger is {\em not maintained} by any cluster and
each cluster maintains only its own {\em view} of the blockchain ledger including
the transactions that access the data shard of the cluster.
The blockchain ledger is indeed the union of all these physical views.

Fig.~\ref{fig:ledger}(b)-(e) show the views of the ledger for
clusters $p_1$, $p_2$, $p_3$, and $p_4$ respectively. 
As can be seen, each cluster $p_i$ maintains only a view of the ledger
consisting of the intra-shard transactions of $p_i$ and
the cross-shard transactions that access $d_i$.
Those transactions are chained together.

\subsection{Transaction Model}

Two main transaction models are used in blockchain systems:
UTXO (Unspent Transaction Output) and Account-based.
In the UTXO model, which is adopted by Bitcoin \cite{nakamoto2008bitcoin} and many other cryptocurrencies,
each transaction spends output from prior transactions and generates new outputs
that can be spent by transactions in the future.
For each transaction in the UTXO model, three conditions need to be satisfied:
first, the sum of the inputs must be equal or greater than the sum of the outputs,
second, every input must be valid and not yet spent, and
third, every input requires a valid signature of its owner.
UTXO provides a higher level of privacy by allowing users to use new addresses for each transaction.

The Account-based model, which is adopted by Ethereum \cite{ethereum17}, is similar to the record keeping in a bank.
The bank tracks how much money each account has, and when users want to spend money,
the bank makes sure that they have enough balance in their account before approving the transaction.
The account-based model is more efficient since the system only
needs to validate that the account has enough balance to pay for the transaction.

UTXO model is used by both OmniLedger \cite{kokoris2018omniledger} and RapidChain \cite{zamani2018rapidchain}
to achieve atomicity of cross-shard transacions without using a distributed commit protocol.
However, as shown in \cite{dang2018towards},
RapidChain fails to achieve isolation and OmniLedger has blocking issues for cross-shard transactions.
\chain, similar to AHL \cite{dang2018towards}, uses the account-based model and performs cross-chain transactions
using a global consensus protocol to achieve correctness.
\section{Consensus with Crash-Only Nodes}\label{sec:cons-crash}

In a permissioned blockchain system,
nodes establish consensus on a unique
order in which entries are appended to the blockchain ledger.
To establish consensus among the nodes, asynchronous fault-tolerant protocols have been used.
Fault-tolerant protocols use the State Machine Replication (SMR) algorithm \cite{lamport1978time}
where nodes agree on an ordering of incoming requests.
The algorithm has to satisfy four main properties \cite{cachin2011introduction}:
(1) {\em agreement:} every correct node must agree on the same value,
(2) {\em Validity (integrity):} if a correct node commits a value, then the value must have been proposed by some correct node,
(3) {\em Consistency (total order):} all correct nodes commit the same value in the same order, and
(4) {\em termination:} eventually every node commits some value.
The first three properties are known as {\em safety} and the termination property is known as {\em liveness}.
Note that consistency (total order) is a trivial property in consensus protocols with a single ordering (cluster),
however, since multiple clusters with different orderings are involved in \chain, consistency  
between different instances of the consensus algorithm needs to be guaranteed.
As shown by Fischer et al. \cite{fischer1985impossibility}, 
in an asynchronous system, where nodes can fail,
consensus has no solution that is both safe and live.
Based on that impossibility result,
similar to most fault-tolerant protocols, in \chain,
safety is ensured in an asynchronous network that can
drop, delay, corrupt, duplicate, or reorder messages.
However, a synchrony assumption is needed to ensure liveness.
In this section, we first show how consensus is established in \chain
for intra-shard and cross-shard transactions in the presence of crash-only nodes.
Then, the primary failure handling routine of \chain is presented and finally the correctness of \chain is proven.

\subsection{Intra-shard consensus}

\begin{figure}[t] \center
\begin{minipage}{.31\textwidth}\centering
\includegraphics[width= \linewidth]{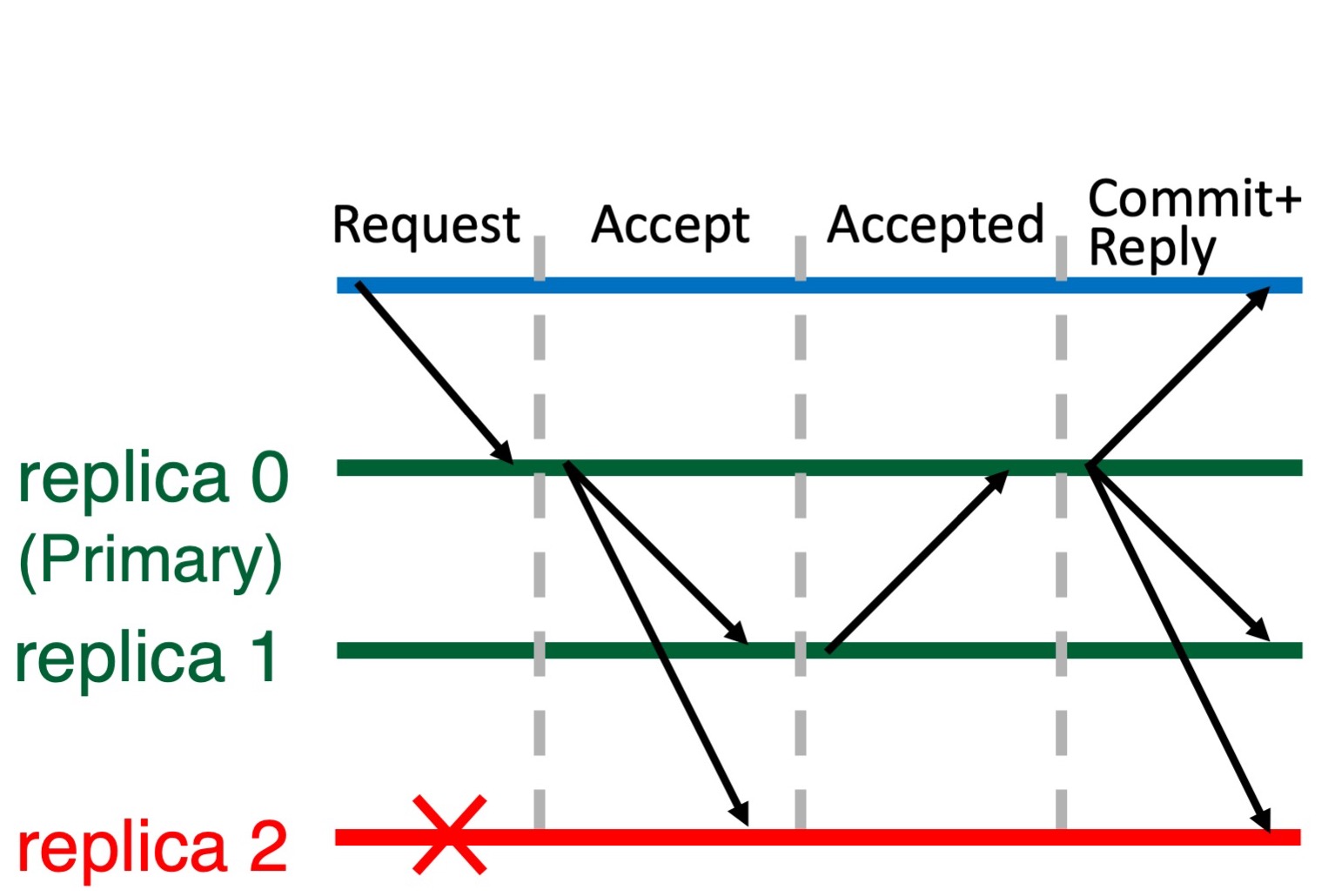}
\end{minipage}
\begin{minipage}{.382\textwidth}\centering
\includegraphics[width= \linewidth]{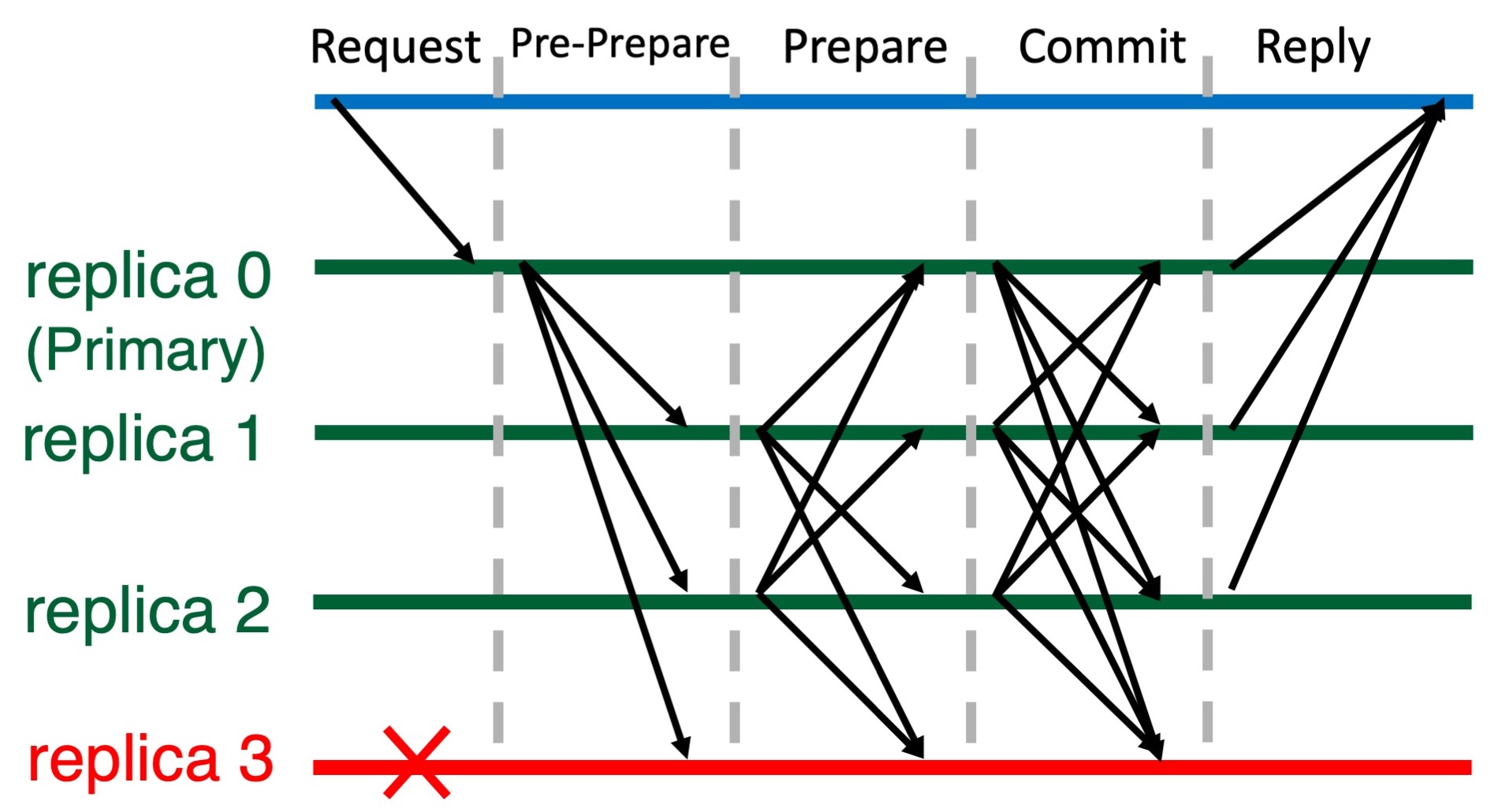}
\end{minipage}
\caption{Normal case operation in (a) Paxos \cite{lamport2001paxos} and (b) PBFT \cite{castro1999practical}}
\label{fig:crash}
\end{figure}

Crash fault-tolerant protocols, e.g., Paxos \cite{lamport2001paxos},
guarantee safety in an asynchronous network using $2f{+}1$ nodes
to overcome the simultaneous crash failure of any $f$ nodes.
\chain uses multi-Paxos, a variation of Paxos,
as can be seen in Figure~\ref{fig:crash}(a),
where {\em the primary} (a pre-elected node that initiates consensus) is relatively stable,
to establish consensus on the order of intra-shard transactions.
In \chain, clients send signed requests to the primary.
The primary then assigns a sequence number to the request (to provide a total order among transactions)
and multicasts a \one message (called {\sf \small accept} in Paxos) 
including the intra-shard transaction to every node within the cluster.
Instead of a sequence number, the primary can also include the cryptographic hash of the previous transaction block, $H(t)$,
in the message where $H(.)$ denotes the cryptographic hash function and
$t$ is the previous block that is ordered by the cluster.
Upon receiving a valid \one message from the primary,
each node sends an \two (i.e., {\sf \small accepted}) message to the primary.
The primary waits for $f$ \two messages from different nodes (plus itself becomes $f+1$),
multicasts a {\em signed} {\sf \small commit} message to every node within the cluster,
appends the transaction block including the transaction and the {\sf \small commit} message
(as evidence of the transaction's validity)
to the blockchain ledger, executes the transaction, 
and sends a {\sf \small reply} to the client.
Upon receiving a {\sf \small commit} message from the primary,
each node appends the transaction block including the transaction and the received {\sf \small commit} message
to its blockchain ledger.
The client also waits for a valid {\sf \small reply} from the primary to accept the result.
If the client does not receive replies soon enough, it multicasts the request
to all nodes within the cluster. If the request has already been processed,
the nodes simply send the execution result back to the client.
Otherwise, if the node is not the primary, it relays the request to the primary.
If the primary does not multicast the request to the nodes of the cluster, it will eventually
be suspected to be faulty by nodes by the nodes.
Note that since {\sf \small commit} messages include the digest (cryptographic hash) of the corresponding transactions,
appending valid signed {\sf \small commit} messages to the blockchain ledger in addition to the transactions,
provides the same level of {\em immutability} guarantee as
including the cryptographic hash of the previous transaction in the transaction block, i.e.,
any attempt to alter the block data can easily be detected.

\subsection{Cross-Shard Consensus}\label{sec:Cross-crash}

\newcommand{\onec}{{\tiny $\langle\text{\ONE}, h_i, d, m \rangle$}\xspace}
\newcommand{\onecc}{{\tiny $\langle\text{\ONEF}{,} h_i{,} d{,} m \rangle$}\xspace}
\newcommand{\twoc}{{\tiny $\langle\text{\TWO}, h_i, h_j, d, r \rangle$}\xspace}
\newcommand{\twocc}{{\tiny $\langle\text{\TWOF}, h_i, h_j, d, r \rangle$}\xspace}
\newcommand{\twoccp}{{\tiny $\langle\text{\TWOF}{,} h_i{,} h_j{,} d{,} \pi(p_j) \rangle$}\xspace}
\newcommand{\threec}{{\tiny $\langle\text{\THREE}, h_i,h_j, ..., h_k, d \rangle_{\sigma_{\pi(p_i)}}$}\xspace}

\begin{algorithm}[t]
\scriptsize
\caption{{\small Cross-shard Consensus with Crash-Only Nodes}}
\label{alg:cross-crash}
\begin{algorithmic}[1]
\State {\bf init():} 
\State \quad $r$ := {\em node\_id}
\State \quad $p_i$ := the cluster that initiates the consensus
\State \quad $\pi(p_j)$ := the primary node of cluster $p_j$
\State \quad $P$ := set of involved clusters
\newline
\State {\bf upon receiving} valid request $m$ and $(r == \pi(p_i))$
\State \quad {\bf multicast} \onec to $P$
\newline
\State {\bf upon receiving} valid \onec from {\em primary} $\pi(p_i)$
\State \quad if $r$ is not waiting for {\scriptsize \sf commit} message of request $m'$
where $m$ and $m'$ intersect in some other cluster $p_k$
\State \qquad {\bf send} \twoc to {\em primary} $\pi(p_i)$
\newline
\State {\bf upon receiving} $f{+}1$ valid matching \twoc from every cluster $p_j$ in $P$ and $(r == \pi(p_i))$
\State \quad {\bf multicast} \threec to $P$
\State \quad {\bf append} the transaction and {\scriptsize \sf commit} message to the ledger
\newline
\State {\bf upon receiving} \threec from $\pi(p_i)$
\State \quad {\bf append} the transaction and {\scriptsize \sf commit} message to the ledger
\end{algorithmic}
\end{algorithm}

Cross-shard transactions access records from different data shards which are maintained by different clusters.
To ensure data consistency, cross-shard transactions have to be appended to
the blockchain ledgers of all involved clusters in the same order.
As a result, consensus among all involved clusters on the order of cross-shard transactions is needed.
In this section, we show how \chain processes cross-shard transactions in a network
consisting of crash-only nodes.

A client sends its request (i.e., a cross-shard transaction) to
the (pre-elected) primary node of a cluster
(i.e., one of the clusters that store data records accessed by the cross-shard transaction).
Note that once a primary node of a cluster is elected,
it initiates all intra-shard transactions of the cluster
as well as cross-shard transactions that are sent to the cluster by clients.
Upon receiving a valid request from a client, primary node $\pi$
initiates the protocol among the involved clusters
by multicasting a \one message
including the transaction to {\em all} nodes of all involved clusters, i.e.,
all clusters that store data records accessed by the cross-shard transaction.
Once a node receives a valid \one message, it sends an \two message to the primary.
The primary waits for matching \two messages from $f{+}1$ nodes of {\em each} involved cluster
to ensure that the majority of every cluster agree with the order of the transaction
(recall that each cluster includes $2f{+}1$ nodes).
The primary then multicasts a \three message to all nodes of every involved cluster,
appends the transaction block to its ledger, executes the transaction,
and sends a \reply to the client.
Upon receiving a \three message from the primary,
each node also appends the transaction block to its ledger. 

Algorithm~\ref{alg:cross-crash} presents the normal case operation for \chain
to process a cross-shard transaction in the presence of crash-only nodes.
Although not explicitly mentioned, every sent and received message is logged by nodes.
As indicated in lines 1-5 of the algorithm,
$p_i$ is the cluster that initiates the transaction,
$\pi(p_j)$ represents the primary node of cluster $p_j$, and
$P$ is the set of involved clusters in the transaction.

As shown in lines 6-7,
upon receiving a valid signed cross-shard request 
$m=\langle\text{\scriptsize \REQ}, op, \tau_c, c\rangle_{\sigma_c}$
from an authorized client $c$ (with timestamp $\tau_c$) to execute operation $op$,
the primary node $\pi(p_i)$ of the initiator cluster $p_i$
assigns sequence number $h_i$ to the request and
multicasts a \one message
$\langle\text{\scriptsize \ONE}, h_i, d, m \rangle$
to the nodes of every involved cluster where
$m$ is the client's request message and
$d=D(m)$ is the digest of $m$.
The sequence number $h_i$ represents the correct order of the transaction block in the initiator cluster $p_i$.
Since all nodes are crash-only, there is no need to sign messages.

Upon receiving a \one message, as indicated in lines 8-10,
each node $r$ of an involved cluster $p_j$
validates the message and its sequence number.
If node $r$ is currently waiting for a \three message of some cross-shard request $m'$ where
the involved clusters of two requests $m$ and $m'$ intersect in some other cluster $p_k$, the node
does not process the new request $m$ before the earlier request $m'$ gets committed.
This ensures that requests are committed in the same order on different clusters.
Otherwise, node $r$ sends an \two message
$\langle\text{\scriptsize \TWO}, h_i, h_j, d, r \rangle$
to primary node $\pi(p_i)$ where $h_j$ is the sequence number, assigned by $r$, that 
represents the correct order of request $m$ in cluster $p_j$ and
$d$ is the digest of $m$.

Once primary $\pi(p_i)$ receives valid matching \two messages
from $f{+}1$ nodes of {\em every} involved cluster $p_j$ with matching $h_j$ and also
$h_i$ and $d$ that match to the \one message sent by $\pi(p_i)$,
as presented in lines 11-13,
it collects all valid sequence numbers (e.g., $h_i$, $h_j$, ..., $h_k$)
from the \two messages of all involved clusters
(e.g., $p_i$, $p_j$, ..., $p_k$) and multicasts a \three message
$\langle\text{\scriptsize \THREE}, [h_i, h_j, ..., h_k], d \rangle_{\sigma_{\pi(p_i)}}$
to the nodes of all involved clusters. The order of sequence numbers $h_i$, $h_j$, ..., $h_k$ 
in the message is an ascending order determined by their cluster ids.
The sequence number, indeed, consists of multiple sub-sequence numbers where
each sub-sequence number presents the local order of the transaction in one of the involved clusters.
The primary signs its \three message because it might be used later by nodes to prove the correctness of the transaction block.

Finally, as shown in lines 14-15, once a node of some cluster $p_j$ receives a valid \three message from primary $\pi(p_i)$,
the node considers the transaction as committed
(even if the node has not sent an \two message for that request).
If all transactions with lower sequence numbers than $h_j$ has been committed,
the node appends the transaction as well as the corresponding \three message to the ledger and executes it.
This ensures that all replicas execute requests
in the same order as required to provide the safety property.

\begin{figure}[t] \center
\includegraphics[width= 0.5 \linewidth]{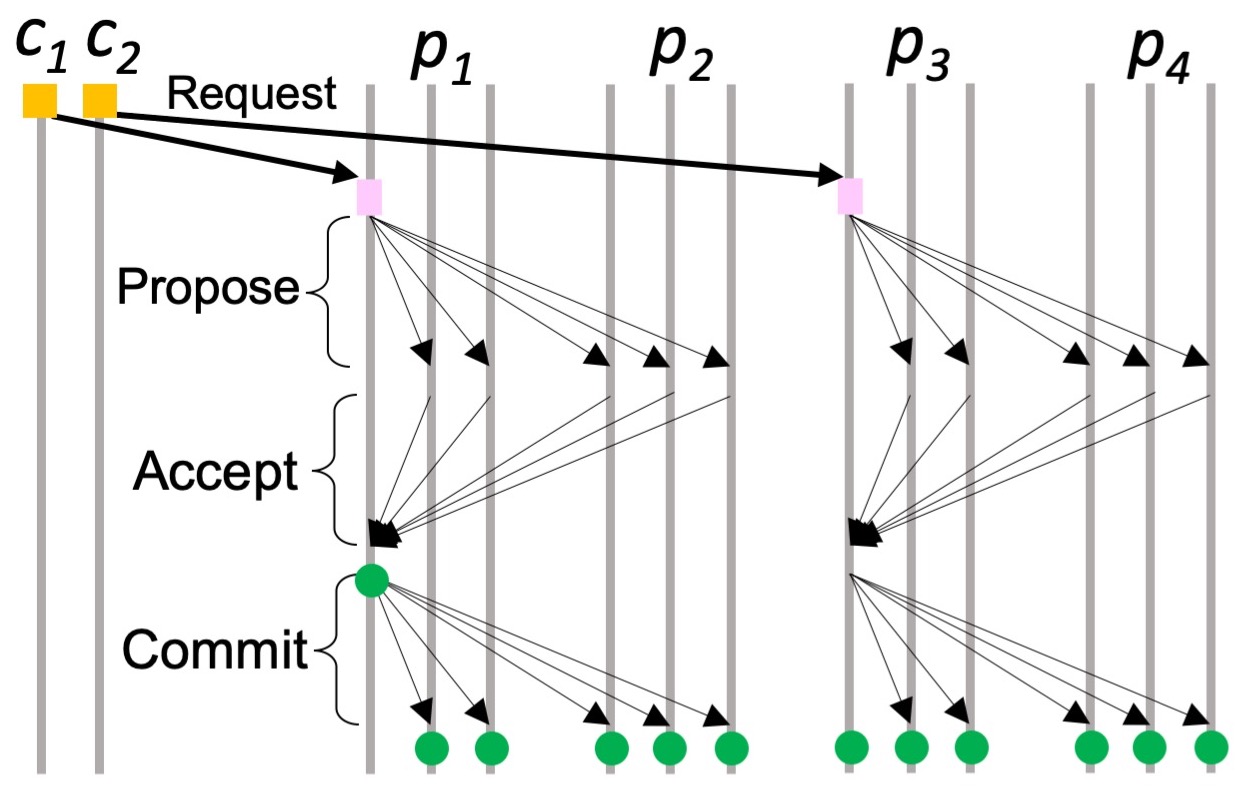}
\caption{Two {\bf concurrent} cross-shard transaction flows for crash-only nodes in \chain where two disjoint
clusters are involved in each transaction}
\label{fig:cross-crash}
\end{figure}

Figure~\ref{fig:cross-crash} shows
the normal case operation for \chain
to execute two concurrent cross-shard transactions in the presence of crash-only nodes where
each transaction accesses two disjoint shards.
The network consists of four clusters where each cluster includes three nodes ($f=1$).

\subsection{Dealing with Conflicting Messages}\label{sec:crash-conf}

In the presented consensus protocol and after multicasting \one messages,
the primary might not receive a quorum of {\em matching} \two messages from $f+1$ nodes of every involved cluster
(i.e., received \two messages might have different sequence numbers) for two reasons.
First, the primary nodes of different clusters might multicast their \one messages in parallel, hence,
different nodes in an overlapping cluster might receive the messages in different order.
Second, nodes might assign inconsistent sequence numbers since they have not necessarily received the latest \one message from
the primary of their own cluster.
We now propose an optimization to reduce the likelihood of such conflicts.
This optimization is demonstrated in Algorithm~\ref{alg:conf-crash}.
In case of non-matching \two messages, as indicated in lines 1-2 of Algorithm~\ref{alg:conf-crash},
primary $\pi(p_i)$ needs to re-initiate the request in {\em only} the {\em conflicting clusters}, i.e.,
clusters that have not sent $f+1$ matching \two messages to the primary node.
However, to prevent any further conflicts, primary $\pi(p_i)$ multicasts a \onef message with the same structure as \one messages
{\em only} to the primary nodes of the conflicting clusters.
Once primary $\pi(p_i)$ sends a \onef message for transaction $m$ to the primary node of a cluster,
$\pi(p_i)$ does not accept any further \two messages for transaction $m$ from that cluster.
As shown in lines 3-4, the primary node of each conflicting cluster
then assigns a sequence number and multicasts a \twof message
(with the same structure as \two messages) to the nodes of its cluster and also the initiator primary $\pi(p_i)$.
Upon receiving a \twof message from the primary of its cluster, as presented in lines 5-6,
each node logs the message and sends a \twof message
with the same sequence number to primary $\pi(p_i)$.
Once primary $\pi(p_i)$ has received valid matching \twof messages
from $f+1$ nodes of {\em every} conflicting cluster, it returns to its normal operation,
as presented in lines 16-18 of Algorithm~\ref{alg:cross-crash},
and multicasts \three messages.

In heavy workloads with a high percentage of cross-shard transaction,
the probability of receiving conflicting \two messages might be high.
Therefore,
instead of multicasting \one messages, waiting for probably conflicting \two messages and then re-initiating the transaction
by multicasting \onef messages,
the primary node of the initiator cluster can initially multicast \onef messages to
the primary nodes of other involved clusters as well as the nodes of its own cluster.
In this way, since the primary of each cluster assigns all sequence numbers for both intra-shard and cross-shard transactions,
no conflicts will occurs.
It should be noted that, this solution comes with an extra (intra-cluster) message passing from
the primary to the nodes of each cluster.
Note that depending on the type of workload and percentage of cross-shard transactions,
\chain can dynamically switch between these two techniques to deal with conflicting messages efficiently.

To deal with conflicting cross-shard transactions,
i.e., cross-shard transactions that are initiated in parallel and overlap in some clusters,
the system designer can also specify {\em mega-primary} nodes.
A mega-primary node is the primary node of one of the clusters in any set $P$ which
initiates all cross-shard transactions that access {\em all} clusters in $P$.
In particular, any transaction that accesses a set of clusters $\{p_i, p_j, ..., p_k\}$
is initiated by the primary node of cluster $i$ where $i = min(i,j, ..., k)$.
For example, if \chain includes three clusters $p_1$, $p_2$, and $p_3$, using a mega primary,
cross-shard transactions that access two clusters $p_1$ and $p_2$, two clusters $p_1$ and $p_3$,
or all three clusters $p_1$, $p_2$, and $p_3$
are initiated by the primary node of $p_1$ (since $1 = min(1,2,3)$) and
cross-shard transactions that access two clusters $p_2$ and $p_3$
are initiated by the primary node of $p_2$ (since $2 = min(2,3)$).
Note that different systems can specify mega primary nodes in different ways
depending on the workload and geographical distance between clusters.

\begin{algorithm}[t]
\scriptsize
\caption{{\small Dealing with Conflicting {\sf \scriptsize ACCEPT} Messages}}
\label{alg:conf-crash}
\begin{algorithmic}[1]
\STATEx ******The configuration is the same as Algorithm~\ref{alg:cross-crash}******
\State if {\sf \scriptsize accept} messages of cluster $p_j$ not matching and $r == \pi(p_i)$
\State \quad {\bf multicast} \onecc to $\pi(p_j)$
\newline
\State {\bf upon receiving} \onecc from $\pi(p_i)$ and  $r {==} \pi(p_j)$
\State \quad {\bf multicast} \twocc to $\pi(p_i)$ and all nodes of $p_j$
\newline
\State {\bf upon receiving} \twoccp from $\pi(p_j)$ and  $r {\in} p_j$
\State \quad {\bf send} \twocc to $\pi(p_i)$
\end{algorithmic}
\end{algorithm}

\subsection{Primary Failure Handling}\label{sec:failure}

The goal of the primary failure handling routine is to provide liveness by allowing
the system to make progress when a primary fails.
It prevents replicas from waiting indefinitely for requests to execute.
The primary failure handling routine must guarantee that it will not introduce any changes in a
history that has been already completed at a correct client.
The routine is triggered by timeout.
When node $r$ of some cluster $p_j$ receives a valid \one message from a primary
for either an intra-shard or a cross-shard transaction,
it starts a timer that expires after some predefined time $\tau$.
Time $\tau$ for cross-shard transactions is larger because
processing cross-shard transactions requires agreement from all involved clusters and takes longer.
If the timer has expired and the node has not received any message from the primary node,
the node suspects that the primary is faulty.
The primary failure handling routine is performed by the nodes of the same cluster as the faulty primary.
However, if a node $r$ is involved in a cross-shard transaction that was
initiated by some other cluster $p_i$ and the timer of $r$ has expired,
node $r$ (of cluster $p_j$) multicasts an \twoq message $\langle\text{\scriptsize \TWOQ}, h_i, h_j, d, r \rangle$ message
to every node of the initiator cluster $p_i$ (the cluster of the faulty primary)
where $h_i$ and $h_j$ are the sequence numbers assigned to the transaction by clusters $p_i$ and $p_j$
(in the corresponding \one and \two messages).
Note that nodes of a cluster do not participate in the primary failure handling routine
of other clusters except for sending \twoq messages.
Otherwise (when the node $r$ and the faulty primary are in the same cluster),
the protocol uses the leader election phase of Paxos \cite{lamport2001paxos} to elect the new primary, and
the new primary will handle all the uncommitted intra- and cross-shard transactions, and take care of the new client requests.
Indeed, similar to Paxos, the node $r$ tries to become the primary node of the cluster
by multicasting a \pre message $\langle\text{\scriptsize \PRE}, H, r \rangle$ to every node of its cluster where
$H$ is a proposal number higher than every sequence number received from the previous primary nodes.
If a node $q$ receives a \pre message with a proposal number $H$ higher than every previous sequence or proposal number received,
the node returns a \pro message $\langle\text{\scriptsize \PRO}, H, q \rangle$ to the sender.
The node that receives $f$ \pro messages (including itself becomes $f+1$) becomes the new primary.

Once the new primary is elected, it multicasts \twoq message $\langle\text{\scriptsize \TWOQ}, h \rangle$
to the nodes of its cluster for any sequence number $h$ ($h <H$) that is still uncommitted (either unknown or accepted).
If $h$ is a cross-shard transaction,
the primary multicasts the \twoq message to every node of all involved clusters.
The primary becomes aware of transaction type (intra- or cross-shard) and hence the involved clusters
either from the received \one message or from the received responses.
Once a node receives an \twoq message for some sequence number $h$,
if the node has already received a \three message (from the previous primary) for sequence number $h$,
it sends a \four message $\langle\langle\text{\scriptsize \FOUR}, h, d \rangle, m\rangle$
to the new primary where $m$ is the \three message received from the previous primary.
Note that for cross-shard transactions, as explained earlier,
$h$ is a combination of several sequence numbers (one per each involved cluster).
Otherwise, if the node has received a \one message for sequence number $h$, sent an \two message
to the previous primary but has not received a \three message, it resends its \two message to the new primary.
Finally, if the node has not received either \one or \three message for sequence number $h_i$,
it sends an \unk message $\langle\text{\scriptsize \UNK}, h, \varnothing \rangle$ to the primary.

The primary collects all responses for each sequence number $h$.
If the primary has received a \three message or $f+1$ matching \one messages
(from each involved cluster in case of a cross-shard transaction)
for a sequence number $h$, it multicasts a \three message to every node (of all involved clusters).
Else, if the primary has received at least one (and at most $f$) matching \one messages
(from any involved cluster in case of cross-shard transactions)
for a sequence number $h$, it multicasts a \one message to every node (of all involved clusters).
Otherwise, the primary multicasts a \one message $\langle\text{\scriptsize \ONE}, h_i, d, \text{no-op} \rangle$
to the backups where the {\sf "no-op"} command leaves the state unchanged.
The last situation happens when the previous primary has assigned sequence number and multicast \one messages to every node,
however, no one has received its message.
Once all unknown transactions to the primary are processed, the primary starts processing new transactions.

\subsection{Correctness Arguments}
Consensus protocols have to satisfy {\em safety} and {\em liveness} properties.
Safety means all correct nodes receive the same requests in the same order whereas
liveness means all correct requests are eventually ordered.
In this section, the safety (agreement, validity, and consistency) and
liveness (termination) properties of \chain in the presence of crash-only nodes are demonstrated.
Since intra-shard transactions follow Paxos, we mainly focus on cross-shard transactions.

\begin{lmm} (\textit{Agreement})
If node $r$ commits request $m$ with sequence number $h$,
no other correct node commits request $m'$ ($m \neq m'$) with the same sequence number $h$.
\end{lmm}

\begin{proof}
Let $m$ and $m'$ ($m \neq m'$) be two committed requests with sequence numbers
$h = [h_i,h_j,h_k,...]$ and $h' = [h'_k,h'_l,h'_m,..]$ respectively.
Committing a request requires
matching \two messages from $f+1$ different nodes of {\em every} involved cluster.
Therefore, if the involved clusters of $m$ and $m'$ intersect in cluster $p_k$,
at least $f+1$ nodes of cluster $p_k$ have sent matching \two messages for $m$, and
similarly, at least $f+1$ nodes of cluster $p_k$ have sent matching \two messages for $m'$.
Since each cluster includes $2f+1$ nodes and nodes are non-malicious, $h_k \neq h'_k$.
Note that the same proof logic applies in special cases where $m$ or $m'$ is an intra-shard transaction
(i.e., $h = h_k$ or $h' = h'_k$).

If the primary fails,
since each committed request has been replicated on a quorum $Q_1$ of $f+1$ nodes and
to become elected primary agreement from a quorum $Q_2$ of $f+1$ nodes is needed,
$Q_1$ and $Q_2$ must intersect in at least one node that is aware of the latest committed request.
Hence, \chain guarantees the agreement property for both intra-shard as well as cross-shard transactions.
\end{proof}

\begin{lmm} (\textit{Validity})
If a correct node $r$ commits $m$, then $m$ must have been proposed by some correct node $\pi$.
\end{lmm}

\begin{proof}
In cross-shard consensus with crash-only nodes, validity is ensured
since crash-only nodes do not send fictitious messages.
\end{proof}

\begin{lmm} (\textit{Consistency})
Let $P_\mu$ denote the set of involved clusters for a request $\mu$.
For any two committed requests $m$ and $m'$ and any two nodes $r_1$ and $r_2$
such that $r_1 \in p_i$, $r_2 \in p_j$, and $\{p_i,p_j\} \in P_m \cap P_{m'}$,
if $m$ is committed before $m'$ in $r_1$, then $m$ is committed before $m'$ in $r_2$.
\end{lmm}

\begin{proof}
As mentioned in Section~\ref{sec:Cross-crash},
once a node $r_1$ of some cluster $p_i$ receives a \one message for some cross-shard transaction $m$,
if the node is involved in some other uncommitted cross-shard transaction $m'$
where ${\mid} P_m \cap P_{m'} {\mid} > 1$, i.e., some other cluster  $p_j$ is also involved in both transactions,
node $r_1$ does not send an \two message for transaction $m$ before $m'$ gets committed.
In this way, since committing request $m$ requires $f+1$ \two messages from {\em every} involved cluster,
$m$ cannot be committed until $m'$ is committed.
As a result the order of committing messages is the same in all involved nodes.
In the special case where $i = j$ (both nodes $r_1$ and $r_2$ belong to the same cluster),
if the primary of the cluster assigns the sequence number,
there will be no inconsistency among nodes.
Otherwise, when the nodes assign sequence numbers and even if
$r_1$ and $r_2$ initially assign inconsistent sequence numbers,
since at least $f+1$ matching \two messages from different nodes of the cluster are needed
to commit a request and the cluster includes $2f+1$ nodes,
the order of committing transactions on nodes $r_1$ and $r_2$ must be consistent.
\end{proof}

\begin{lmm}(\textit{Termination})
A request $m$ issued by a correct client eventually completes.
\end{lmm}

\begin{proof}
\chain, as mentioned earlier and due to the FLP impossibility result \cite{fischer1985impossibility},
guarantees liveness only during periods of synchrony.
To show that a request issued by a correct client eventually completes, we need to address three scenarios.
First, if the primary is non-faulty and \two messages are non-conflicting, as shown in Algorithm~\ref{alg:cross-crash},
the protocol ensures that the correct client receives \reply from the primary.
Second, if a non-faulty primary has multicast \one messages but not received matching \two messages
from $f+1$ nodes of every involved clusters, as explained in Sections~\ref{sec:crash-conf}, the primary
re-initiates the transaction by multicasting \onef messages to only the primary nodes of the involved clusters.
In this way, since the primary node of each cluster assigns the sequence number (in its \twof message),
\twof messages that are received from each cluster are matching, thus increasing the chances of termination.
If the primary node of any involved cluster has failed before multicasting \twof messages,
the primary failure handling routine will trigger and the new elected primary will handle the transaction.
Third, if the primary fails, as explained in Sections~\ref{sec:failure},
the nodes that are involved in an uncommitted transaction (initiated by the faulty primary)
detect its failure (using timeouts) and trigger the primary failure handling.
The new primary then will handle all uncommitted transactions.
\end{proof}
\section{Consensus with Byzantine Nodes}\label{sec:cons-byz}

In this section, intra- and cross-shard consensus
in the presence of Byzantine nodes are  presented first followed by the view change routine.
Then, the correctness of \chain with malicious failures is proven, and finally,
an optimization for clustered networks is discussed,

\subsection{Intra-Shard Consensus}

Most Byzantine fault-tolerant protocols, e.g., PBFT \cite{castro1999practical}, require $3f{+}1$ nodes
to guarantee safety in the presence of at most $f$ {\em malicious} nodes.
PBFT consists of {\em agreement} and {\em view change} routines where
the agreement routine orders requests for execution by the nodes, and 
the view change routine coordinates the election of a new primary when
the current primary is faulty.
The nodes move through a succession of configurations
called {\em views} \cite{el1985efficient}\cite{el1985availability} where
in each view, one node, called {\em the primary}, initiates the protocol
and the others are {\em backups}.

To establish consensus on the order of intra-shard transactions
during a normal case execution of PBFT,
as can be seen in Figure~\ref{fig:crash}(b),
a client $c$ requests an intra-shard transaction
by sending a signed {\sf \small request} message
to the primary.
When the primary receives a valid request from an authorized client,
it initiates the consensus protocol
by assigning a sequence number and multicasting a \one (called {\sf \small pre-prepare} in original PBFT)
message including the requested transaction to all nodes within the cluster.
Once a node receives a valid \one message from the primary, it multicasts
an \two ({\sf \small prepare}) message to every node within the cluster.
Each node then waits for $2f$ valid \two messages from different nodes (including itself)
that match the \one message
and then multicasts a {\sf \small commit} message to all the nodes
within the cluster.
Once a node receives $2f$ valid {\sf \small commit} messages from different nodes
that match its own {\sf \small commit} message,
it appends the transaction block including all $2f+1$ {\sf \small commit} message to the ledger (to ensure immutability),
executes the transaction, and sends a \reply to the client.
Finally, the client waits for $f+1$ valid matching responses from different replicas
to make sure at least one correct replica executed its request.
If the client does not receive \reply messages soon enough, it multicasts the request
to all nodes within the cluster. If the request has already been processed,
the nodes simply re-send the \reply message to the client
(replicas remember the last reply message they sent to each client).
Otherwise, if the node is not the primary, it relays the request to the primary.
If the primary does not multicast the request to the nodes of the cluster, it will eventually
be suspected to be faulty by nodes to cause a view change.

\subsection{Cross-Shard Consensus with Byzantine Nodes}\label{sec:Cross-byzantine}

\newcommand{\oneip}{{\tiny $\langle\langle\text{\ONE}, h_i, v_i d \rangle_{\sigma_{\pi(p_i)}}, m \rangle$}\xspace}
\newcommand{\twop}{{\tiny $\langle\text{\TWO}, h_i, h_j, v_i, v_j, d, r \rangle_{\sigma_r}$}\xspace}
\newcommand{\twopp}{{\tiny $\langle\text{\TWOF}, h_i{,} h_j, v_i, v_j, d, r \rangle_{\sigma_r}$}\xspace}
\newcommand{\twoppj}{{\tiny $\langle\text{\TWOF}, h_i{,} h_j{,} v_i{,} v_j{,} d{,} \pi(p_j) \rangle_{\sigma_{\pi(p_j)}}$}\xspace}
\newcommand{\threep}{{\tiny $\langle\text{\THREE}, h_i, h_j, ..., h_k, v_i, v_j, ..., v_k, d, r \rangle_{\sigma_r}$}\xspace}

\begin{algorithm}[t]
\scriptsize
\caption{{\small Cross-shard Consensus with Byzantine Nodes}}
\label{alg:cross-Byzantine}
\begin{algorithmic}[1]
\State {\bf init():} 
\State \quad $r$ := {\em node\_id}
\State \quad $p_i$ := the cluster that initiates the consensus
\State \quad $\pi(p_j)$ := the primary node of $p_j$
\State \quad $P$ := set of involved clusters
\newline
\State {\bf upon receiving} valid transaction $m$ and $(r == \pi(p_i))$
\State \quad {\bf multicast} \oneip to $P$
\newline
\State {\bf upon receiving} valid \oneip from $\pi(p_i)$
\State \quad if $r$ is not involved in any uncommitted request $m'$
where $m$ and $m'$ intersect in some other cluster $p_k$
\State \qquad {\bf multicast} \twop to $P$
\newline
\State {\bf upon receiving} valid matching \twop from $2f{+}1$ different nodes of every cluster $p_j$ in $P$
\State \quad {\bf multicast} \threep to $P$
\newline
\State {\bf upon receiving} valid \threep from $2f+1$ nodes of every cluster in $P$
\State \quad {\bf append} the transaction and {\scriptsize \sf commit} messages to the ledger
\end{algorithmic}
\end{algorithm}

In the presence of malicious nodes, a Byzantine fault-tolerant protocol is needed
to order cross-shard transactions where for each cross-shard transaction, similar to the crash-only case,
agreement from all involved clusters is needed.
Unlike in the case of crash failure where the quorum size is $f+1$,
in consensus with Byzantine nodes, the quorum size is $2f+1$.
In addition and due to the potential malicious behaviour of the primary node,
all nodes of every involved cluster multicast both \two and \three messages to each other.
In cross-shard consensus with Byzantine node, similar to PBFT, 
nodes of each cluster move through views where views are numbered consecutively.
Node $\pi$ ($1 \leq \pi \leq 3f{+}1$) is {\em the primary} of view $v$
if $\pi {=} (v \mod (3f{+}1))$.

In the presence of malicious nodes, and upon receiving a valid request (cross-shard transaction) from a client,
similar to the crash-only case,
primary node $\pi$ initiates the protocol among the involved clusters
by multicasting a \one message including the transaction to {\em all} nodes of all involved clusters.
Once a node receives a valid \one message, it multicasts an \two message to all nodes of every involved clusters.
Each node then waits for $2f+1$ matching valid \two messages from different nodes of {\em each} involved cluster before
multicasting a \three message to all nodes of the involved clusters.
Upon receiving $2f+1$ matching valid \three message from different nodes of {\em each} involved cluster,
each node appends the transaction block to the ledger.

The normal case operation for \chain
to process a cross-shard transaction in the presence of Byzantine nodes
is presented in Algorithm~\ref{alg:cross-Byzantine}.
Similar to Algorithm~\ref{alg:cross-crash} and as shown in lines 1-5,
$p_i$ is the initiator cluster,
$P$ is the set of involved clusters, and
$\pi(p_j)$ indicates the primary node of cluster $p_j$.

Once the initiator primary $\pi(p_i)$ receives a valid signed cross-shard request
from an authorized client,
as presented in lines 6-7,
the primary assigns sequence number $h_i$ to the request and
multicasts a {\em signed} \one message including
sequence number $h_i$, view number $v_i$
(that indicates the view of cluster $p_i$ in which the message is being sent)
and digest $d$ of the request.
As before, sequence number $h_i$ is used
to ensure that the new transaction block is ordered correctly with respect to
the blocks that the cluster has been involved in.
Requests are piggybacked in \one messages
to keep \one messages small.
Since the network might include malicious nodes,
the primary signs its message.

Once a node $r$ of an involved cluster $p_j$ receives a \one message for a request $m$, as indicated in lines 8-10,
it validates the signature and message digest.
If the node belongs to the initiator cluster ($i=j$), it also checks $h_i$ to be valid (within a certain range)
since a malicious primary might multicast a request with an invalid sequence number.
Furthermore, if the node is currently involved in an uncommitted cross-shard request $m'$ where
the involved clusters of two requests $m$ and $m'$ overlap in some other cluster, 
as explained in the crash-only case, the node
does not process the new request $m$ before the earlier request $m'$ is processed.
This is needed to ensure requests are committed in the same order on different clusters.
The node then multicasts a signed \two message including
the corresponding sequence number $h_j$ (that represents the order of request $m$ in cluster $p_j$),
the view number $v_j$ of cluster $p_j$ as well as the digest $d$ of request $m$ 
to {\em every} node of {\em all} involved clusters.

Each node waits for valid \two messages with matching sequence and view numbers
from $2f{+}1$ nodes of {\em every} involved cluster
with $h_i$, and $d$ that match the \one message which is sent by primary $\pi(p_i)$.
We define the predicate {\sf accepted-local}$_{p_j}(m, h_i, h_j, v_i,v_j,r)$ to be true if and only if node $r$
has received the request $m$, a \one for $m$ with sequence number $h_i$ in view $v_i$ of the initiator cluster $p_i$
and $2f+1$ singed \two messages form different nodes of an involved cluster $p_j$ that match the \one message.
The predicate {\sf accepted}$(m, h, v, r)$ where
$h = [h_i, h_j, ..., h_k]$ and $v = [v_i, v_j, ..., v_k]$
is then defined to be true on node $r$
if and only if {\sf accepted-local}$_{p_j}$ is true for {\em every} involved cluster $p_j$ in cross-shard request $m$.
The order of sequence numbers and view numbers in the predicate is an ascending order determined by their cluster ids.
Here, since nodes might behave maliciously, each cluster includes $3f+1$ nodes and
$2f+1$ matching messages from {\em all} involved clusters for each step of the protocol are needed.
The \one and \two phases of the algorithm basically guarantee that non-faulty nodes agree on a total order
for the transactions.
When {\sf accepted}$(m, h, v, r)$ becomes true,
as presented in lines 11-12, the node $r$ multicasts a signed \three message
$\langle\text{\scriptsize \THREE}, h, v, d, r \rangle_{\sigma_r}$
to every node of all involved clusters.

Finally, as shown in lines 13-14, each node waits for valid matching signed \three messages 
from $2f+1$ nodes of {\em every} involved clusters that match its \three message.
The predicate {\sf committed-local}$_{p_j}(m, h, v, r)$
is defined to be true on node $r$ if and only if 
{\sf accepted}$(m, h, v, r)$ is true and node $r$ has accepted $2f+1$ valid matching \three
messages from different nodes of cluster $p_j$ that match the \one message for cross-shard request $m$.
The predicate {\sf committed}$(m, h, v, r)$ is then defined to be true on node $r$
if and only if {\sf committed-local}$_{p_j}$ is true for {\em every} involved cluster $p_j$ in cross-shard request $m$.
The {\sf committed} predicate indeed shows that at least $f+1$ nodes of each involved cluster have multicast valid \three messages.
When the {\sf committed} predicate becomes true, the node considers the transaction as committed.
If the node has executed all transactions with lower sequence numbers than $h_j$,
it appends the transaction as well as the corresponding \three message to the ledger and executes it.

\begin{figure}[t] \center
\includegraphics[width= 0.5 \linewidth]{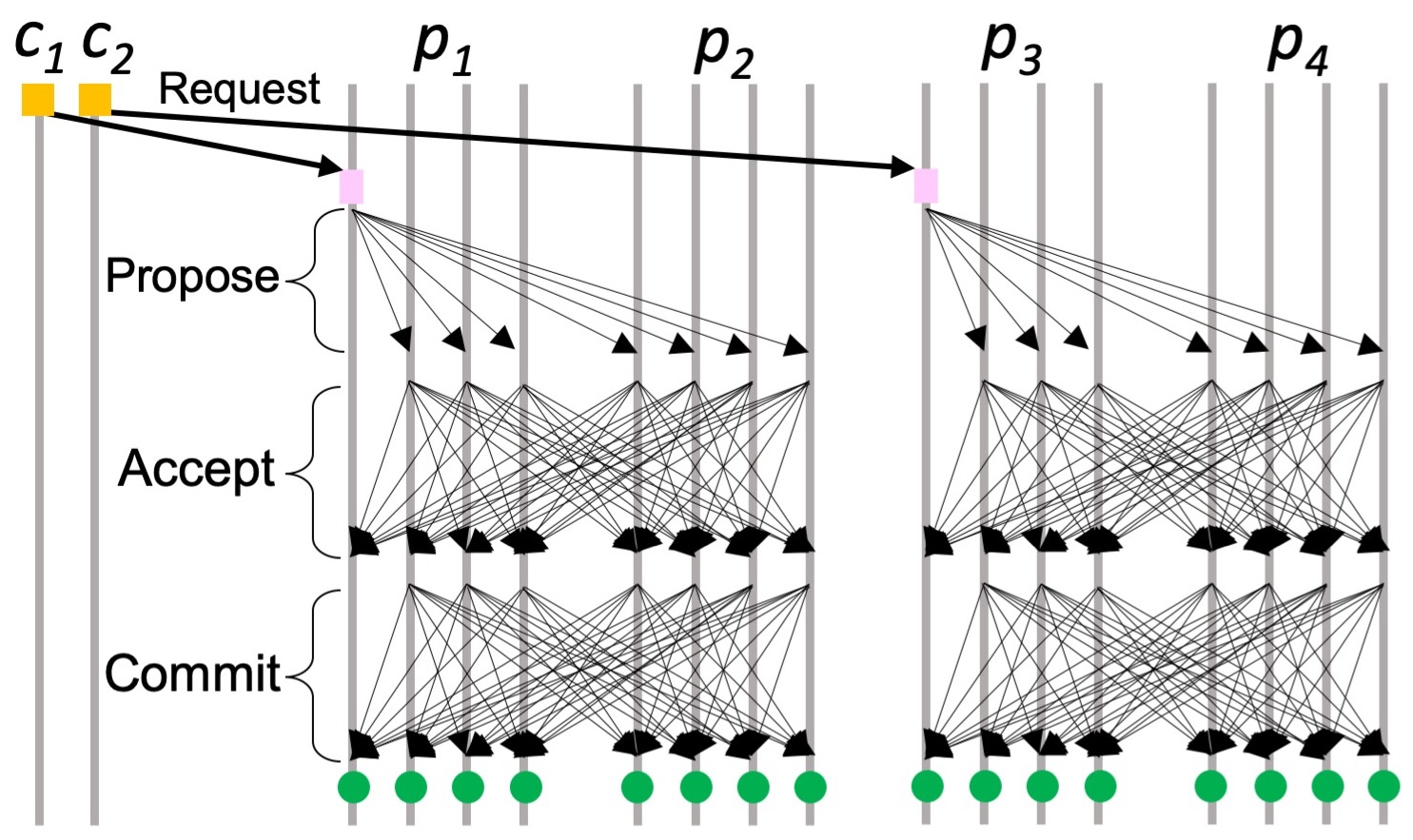}
\caption{Two {\bf concurrent} cross-shard transaction flows for Byzantine nodes in \chain}
\label{fig:cross}
\end{figure}

Figure~\ref{fig:cross} shows the processing of two concurrent cross-shard transactions in
the presence of Byzantine nodes where each transaction accesses two disjoint data shards.
The network consists of four clusters where each cluster includes four nodes ($f=1$).

\subsection{Dealing with Conflicting Messages}\label{sec:byz-conf}

\begin{algorithm}[H]
\scriptsize
\caption{{\small Dealing with Conflicting {\sf \scriptsize ACCEPT} Messages}}
\label{alg:conf-byz}
\begin{algorithmic}[1]
\STATEx ******The configuration is the same as Algorithm~\ref{alg:cross-Byzantine}******
\State if {\sf \scriptsize accept} messages of cluster $p_j$ not matching and $(r == \pi(p_j))$
\State \quad {\bf multicast} \twopp to nodes of $p_j$
\newline
\State {\bf upon receiving} \twoppj and  $r {\in} p_j$
\State \quad if less than $2f+1$ valid {\sf \scriptsize accept} messages from $p_j$ for $m$ is logged
\State \qquad {\bf multicast} \twopp to $P$
\end{algorithmic}
\end{algorithm}

In the consensus protocol with Byzantine nodes, similar to the crash-only case,
nodes might not receive a quorum of $2f+1$ {\em matching} \two messages from every cluster
due to conflicting \two messages.
In such a situation, as presented in lines 1-2 of Algorithm~\ref{alg:conf-byz},
the primary node of each conflicting cluster $p_j$ (i.e., a cluster with less than $2f+1$ matching \two messages)
multicasts a \twof message (with the same structure as \two messages) to the nodes of its own cluster.
Once a node receives a \twof message for some cross-shard transaction $m$ from the primary node of its cluster,
as shown in lines 3-5, it first validates the message.
If the node has already received $2f+1$ matching \two messages for transaction $m$ from the nodes of its cluster
(which might happen in case of a malicious primary), the node simply ignores the received \twof message.
Otherwise, the node multicasts a \twof message to all nodes of every involved cluster.

In heavy workloads with a high percentage of cross-shard transactions,
since the probability of receiving conflicting \two messages is high, Similar to the crash-only case, 
the primary node of the initiator cluster can initially multicast \onef messages to
the primary nodes of other involved clusters as well as the nodes of its own cluster.

In addition, \chain can employ the mega-primary technique where
for any set $P$ of clusters, the primary node of one of the clusters, called {\em mega-primary},
initiates all cross-transactions that access {\em all} clusters in $P$.

\subsection{View Change}\label{sec:viewchange}

The view change routine provides liveness by allowing the system to make progress when a primary fails.
Similar to the crash-only case, view changes are triggered by timeout.
If the timer of some node $r$ expires, node $r$ suspects that the primary is faulty and starts a view change.
There are two cases.
First, if the initiator primary is not in the cluster of node $r$, similar to the crash-only case,
node $r$ multicasts a signed \twoq message
to every node of the initiator cluster (the cluster of the faulty primary).
If a node receives \twoq messages from $2f+1$ different nodes of another cluster,
the node suspects that the primary of its cluster is faulty and initiates a view change.
Second, when node $r$ and the faulty primary are in the same cluster,
similar to PBFT, node $r$ initiates a view change.
To begin the view change routine, node $r$ stops accepting \one, \two, \twof, and \three messages and
multicasts a \vchange message 
$\big\langle\text{\scriptsize \VCHANGE}, v+1, h, \xi, {\cal A}, {\cal C}, r \big\rangle_{\sigma_r}$
to every node within its cluster where
$h$ is the sequence number of the last stable checkpoint (is explained later) known to $r$,
$\xi$ is the proof of checkpoint,
$\cal A$ is the set of received valid intra- and cross-shard \two and \twof messages, and
$\cal C$ is the set of received valid \three messages
for requests with a sequence number higher than $h$.
An \twoq message is valid if it is received from at least $2f+1$ different nodes of the same cluster.
Note that, \chain, similar to PBFT, use the state transfer technique to checkpoint the state of different nodes. 
Each node generates checkpoints periodically when a request
sequence number is divisible by some constant (checkpoint period) and multicasts them to other nodes in its cluster.
Once a node has received $2f+1$ checkpoint messages (called the proof of checkpoint) for a sequence number,
the checkpoint becomes stable.

When primary $\pi'(p_j)$ of new view $v+1$ receives
$2f$ valid \vchange messages
from different nodes of its cluster, it multicasts a
$\big\langle\text{\scriptsize \NEWV}, v{+}1, \Sigma, {\cal P'}, {\cal C'},\big\rangle_{\sigma_{\pi'(p_j)}}$
message to all nodes where
$\Sigma$ is the set of $2f+1$ valid \vchange messages
($2f$ messages from other nodes plus its own message), and
$\cal P'$ and $\cal C'$ are two sets of \one and \three messages respectively
which are constructed as follows.

Let $l$ be the sequence number of the latest checkpoint, and
$h$ be the highest sequence number of a \one message in all the received $\cal A$ sets.
For each sequence number $n$ where $l < n \leq h$.

$\bullet$ It first checks all \three messages in set $\cal C$ of the received \vchange messages.
If the primary finds $2f+1$ valid matching \three messages from either cluster $p_j$ for intra-shard request $m$ or
from every cluster for cross-shard request $m$,
the primary adds the \three messages to $\cal C'$.

$\bullet$ If the primary node $\pi'(p_j)$ finds a set of $2f+1$ matching valid \two messages for an intra-shard transaction or
a set of $2f+1$ matching valid \two or \twof messages coming from the same cluster for a cross-shard transaction,
the primary $\pi'(p_j)$ adds a
$\big\langle\text{\scriptsize \ONE}, v{+}1, n, d \big\rangle_{\sigma_{\pi'(p_j)}}$
to $\cal P'$ where $d$ is the digest of the request.

$\bullet$ If \two messages of the nodes of its cluster are not matching and
the request is a cross-shard transaction initiated by other cluster,
the primary assigns a sequence number and adds 
$\big\langle\text{\scriptsize \TWOF}, v{+}1, n, d \big\rangle_{\sigma_{\pi'(p_j)}}$
to $\cal P'$ where $d$ is the digest of the request.

$\bullet$ Otherwise, the primary adds a
$\big\langle\text{\scriptsize \ONE}, v{+}1, n, d^{\emptyset} \big\rangle_{\sigma_{\pi'(p_j)}}$
to $\cal P'$ where $d^{\emptyset}$ is the digest of a special no-op command
that is transmitted by the protocol like other requests but leaves the state unchanged.

The primary then inserts all the messages in $\cal P'$ and $\cal C'$ to its log.
It also checks the log to make sure its log contains the latest stable checkpoint.
If not, the primary inserts \checkp messages for the checkpoint $l$ and obtain missing blocks
in its blockchain form another node.
For each cross-shard transaction, the primary also multicasts the corresponding message, e.g., \one, \two, or \twof
to the nodes of all involved clusters.

Once a node in view $v$ receives a valid \newv message from the primary of view $v+1$,
the node logs all messages,
updates its checkpoint in the same way as the primary, and
for each \one or \twof message, multicasts an \two or \twof message (respectively) to the nodes of the involved clusters.
Note that non-faulty nodes in view $v$ will not accept a \one message for a new view $v' > v$
without having received a \newv message for $v'$.

Note that nodes redo the protocol for requests with sequence number between $l$ and $h$, however,
they do not re-execute requests. In addition, if a node does not have a request message or 
or a stable checkpoint, it obtains missing information from another node.

\subsection{Correctness Arguments}
In this section we demonstrate how \chain satisfies safety (agreement, validity, and consistency) and
liveness (termination) properties in the presence of Byzantine nodes.

\begin{lmm} (\textit{Agreement})
If node $r$ commits request $m$ with sequence number $h$,
no other correct node commits request $m'$ ($m \neq m'$) with the same sequence number $h$.
\end{lmm}

\begin{proof}
The \one and \two phases of the Byzantine cross-shard consensus protocol
guarantee that correct nodes agree on a total order of
requests within a view.
Indeed, if the {\sf accepted}$(m, h, v, r)$ predicate where $h = [h_i, h_j, ..., h_k]$ and
$v=[v_i, v_j, ..., v_k]$ is true, 
then {\sf accepted}$(m', h, v, q)$ is false for any non-faulty
node $q$ (including $r=q$) and any $m'$ such that $m \neq m'$.
This is true because $(m, h, v, r)$
implies that {\sf accepted-local}$_{p_j}(m, h_i, h_j, v_i,v_j,r)$ is true for each involved cluster $p_j$ and
since each cluster include $3f+1$ nodes, at least $2f+1$ nodes within the cluster (from which at least $f+1$ nodes are non-faulty)
have sent \two (or \one) messages
for request $m$ with sequence number $h_j$ in view $v_j$.
As a result, for {\sf accepted}$(m', h, v, q)$ to be true, at least one
of those non-faulty nodes needs to have sent two conflicting \two messages
with the same sequence number, same view number, but different message digest.
This condition guarantees that first, a malicious primary cannot violate the safety and
second, at most one of the concurrent {\em conflicting} transactions, i.e.,
transactions that overlap in at least one cluster,
can collect the required number of messages ($2f+1$) from each overlapping cluster.

Across different views, the view-change routine of \chain guarantees that non-faulty nodes of some cluster $p_j$
agree on the sequence number of requests that are {\sf committed-local} in different views at different node.
The {\sf committed-local}$_{p_j}$ predicate becomes correct on node $r$ if
$r$ has received a quorum $Q_1$ of $2f+1$ matching \three messages from different nodes of cluster $p_j$.
To change the view of cluster $p_j$, a quorum $Q_2$ of $2f+1$ valid \vchange messages is needed.
Since there are $3f+1$ nodes in each cluster, $Q_1$ and $Q_2$ intersect in at least one correct node, thus
if a request is {\sf accepted} in a previous view, it is propagated to subsequent views.
\end{proof}

\begin{lmm} (\textit{Validity})
If a correct node $r$ commits $m$, then $m$ must have been proposed by some correct node $\pi$.
\end{lmm}

\begin{proof}
In the presence of Byzantine nodes, validity is guaranteed mainly based on
standard cryptographic assumptions about collision-resistant hashes, encryption, and signatures
which the adversary cannot subvert them (as explained in Section~\ref{sec:model}).
Since the request as well as all messages are signed and
either the request or its digest is included in each message
(to prevent changes and alterations to any part of the message), and
in each step $2f+1$ matching messages (from each cluster) are required,
if a request is committed, the same request must have been proposed earlier.
\end{proof}

\begin{lmm} (\textit{Consistency})
Let $P_\mu$ denote the set of involved clusters for a request $\mu$.
For any two committed requests $m$ and $m'$ and any two nodes $r_1$ and $r_2$
such that $r_1 \in p_i$, $r_2 \in p_j$, and $\{p_i,p_j\} \in P_m \cap P_{m'}$,
if $m$ is committed before $m'$ in $r_1$, then $m$ is committed before $m'$ in $r_2$.
\end{lmm}

\begin{proof}
Consistency is guaranteed in the same way as crash-only nodes (lemma 3.3).
\end{proof}

\begin{lmm}(\textit{Termination})
A request $m$ issued by a correct client eventually completes.
\end{lmm}

\begin{proof}
To provide termination during periods of synchrony, similar to the crash-only case,
three scenarios need to be addressed.
If the primary in non-faulty and \two messages are non-conflicting, following Algorithm~\ref{alg:cross-Byzantine},
request $m$ completes.
If the primary in non-faulty, however \two messages are conflicting, as mentioned in Section~\ref{sec:byz-conf},
the request will be re-initiated in the conflicting clusters using \onef messages.
Finally, view change routines (Section~\ref{sec:viewchange}) handle primary failures.
\end{proof}

\subsection{An Optimization for Clustered Networks}

We now illustrate
how prior knowledge of where the faulty nodes are placed could help in increasing the number of clusters,
and hence parallelism and overall performance.

In \chain and in the presence of crash-only nodes,
we assume that the number of nodes is much more than $2f+1$ and therefore, partition
the network into clusters of $2f+1$ nodes.
This is needed because we are not aware of {\em where} the $f$ faulty nodes are placed.
As a result, since they all might be in the same cluster, to guarantee safety each cluster includes
$2f+1$ nodes.
Similarly, and in the presence of Byzantine nodes, each cluster consists of $3f+1$ nodes.
However, if we have some prior knowledge of where the faulty nodes are placed,
we might be able to increase the number of clusters.
In particular, nodes might be (geographically) partitioned into several groups (e.g., clouds)
where $f$ is known for each individual group of nodes.
Hence, clustering can be performed within each group instead of the entire network.
Indeed, different cloud environments might have different failure properties, e.g.,
while renting nodes from a particular cloud might be expensive,
the maximum number of possible concurrent failures, $f$, in that cloud could be smaller than a cloud with cheaper nodes.
As an example, consider a network of Byzantine nodes with $n=23$ and $f=3$ where
nodes are partitioned into two groups of $A$ and $B$
(placed in two different cloud environments) such that
$n_A =7$, $n_B = 16$, $f_A= 2$, and $f_B =1$.
Without being aware of $A$ and $B$, since there are totally $23$ nodes and $f=3$,
the number of clusters is $|P| = \frac{n}{3f+1} = \frac{23}{10} =2$.
However, knowing $f_A$ and $f_B$, we can cluster $A$ and $B$ separately and as a result,
$|P_A| = \frac{n_A}{3f_A+1} = \frac{7}{7} =1$ and $|P_B| = \frac{n_B}{3f_B+1} = \frac{16}{4} =4$.
Thus, the network is partitioned into five clusters (three more clusters in comparison to the previous case).
This is useful especially in cloud environments where
nodes are placed in different clouds with different properties (e.g., different $f$).
Note that the same technique can be applied when the nodes are crash-only.

Furthermore, \chain can be extended to support hybrid cloud environments where
clusters (clouds) have different failure models, e.g., private clouds with crash-only nodes and
public cloud with malicious nodes.
In such a setting, different clusters, depending on the failure model of their nodes,
might use different consensus protocols, i.e., crash or Byzantine fault-tolerant protocols, and
a hybrid fault-tolerant protocol like SeeMoRe \cite{amiri2019seemore} can be used to order
cross-shard transactions.

\section{Experimental Evaluations}\label{sec:eval}

In this section, we conduct several experiments to evaluate \chain.
We have implemented a blockchain-based accounting application where the data records are
client accounts (every client might have several accounts).
Clients of the application can initiate transactions to
transfer assets from one or more of their accounts to other accounts
(accounts might be in the same shard or different shards).
A transaction might read and write several records.
The experiments were conducted on the Amazon EC2 platform.
Each VM is c4.2xlarge instance with 8 vCPUs and 15GB RAM,
Intel Xeon E5-2666 v3 processor clocked at 3.50 GHz.
When reporting throughput measurements, we use an increasing
number of clients running on a single VM,
until the end-to-end throughput is saturated,
and state the throughput ($x$ axis) and latency ($y$ axis) just below saturation.
Throughput and latency numbers are reported as the average measured during the steady
state of an experiment.

\begin{figure*}[t!]
\Large
\begin{minipage}{.25\textwidth}\centering
\begin{tikzpicture}[scale=0.47]
\begin{axis}[
    xlabel={Throughput [ktrans/sec]},
    ylabel={Latency [ms]},
    xmin=0, xmax=40,
    ymin=0, ymax=200,
    xtick={0,10,20,30},
    ytick={50,100,150},
    legend style={at={(axis cs:15,190)},anchor=north west}, 
    ymajorgrids=true,
    grid style=dashed,
]

\addplot[
    color=red,
    mark=o,
    mark size=4pt,
    line width=0.5mm,
    ]
    coordinates {
    (0.304,8)(2.214,22)(4.413,36)(6.583,61)(8.783,94)(9.104,164)};
    
\addplot[
    color=green,
    mark=square,
    mark size=4pt,
    line width=0.5mm,
    ]
    coordinates {
    (0.304,6)(2.198,11)(4.374,18)(6.491,37)(8.711,58)(10.683,75)(10.934,144)};

\addplot[
    color=blue,
    mark=*,
    mark size=4pt,
    line width=0.5mm,
    ]
    coordinates {
    (0.099,7)(0.980,9)(8.899,23)(17.750,37)(26.550,59)(35.230,91)(38.210,180)};
    
\addplot[
    color=black,
    mark=triangle,
    mark size=4pt,
    line width=0.5mm,
    ]
    coordinates {
    (0.099,7)(0.980,9)(8.899,23)(17.750,37)(26.550,59)(35.230,91)(38.210,180)};


\addlegendentry{APR-C}
\addlegendentry{F-Paxos}
\addlegendentry{SharPer}
\addlegendentry{AHL-C}

\end{axis}
\end{tikzpicture}
{\footnotesize (a) $0\%$ Cross-shard}
\end{minipage}\hfill
\begin{minipage}{.25\textwidth} \centering
\begin{tikzpicture}[scale=0.47]
\begin{axis}[
    xlabel={Throughput [ktrans/sec]},
    ylabel={Latency [ms]},
    xmin=0, xmax=25,
    ymin=0, ymax=200,
    xtick={0,5,10,15,20},
    ytick={50,100,150},
    legend style={at={(axis cs:11,115)},anchor=south west}, 
    ymajorgrids=true,
    grid style=dashed,
]

\addplot[
    color=red,
    mark=o,
    mark size=4pt,
    line width=0.5mm,
    ]
    coordinates {
    (0.304,8)(2.214,22)(4.413,36)(6.583,61)(8.783,94)(9.104,164)};
    
\addplot[
    color=green,
    mark=square,
    mark size=4pt,
    line width=0.5mm,
    ]
    coordinates {
    (0.304,6)(2.198,11)(4.374,18)(6.491,37)(8.711,58)(10.483,75)(10.534,144)};

\addplot[
    color=blue,
    mark=*,
    mark size=4pt,
    line width=0.5mm,
    ]
    coordinates {
  (0.888,3)(4.875,11)(10.105,31)(17.142,61)(22.950,104)(23.716,173)};

\addplot[
    color=black,
    mark=triangle,
    mark size=4pt,
    line width=0.5mm,
    ]
    coordinates {
  (0.804,5)(4.229,13)(9.172,35)(15.413,65)(21.443,113)(21.692,164)};

\addlegendentry{APR-C}
\addlegendentry{F-Paxos}
\addlegendentry{SharPer}
\addlegendentry{AHL-C}
 
\end{axis}
\end{tikzpicture}
{\footnotesize (b) $20\%$ Cross-shard}
\end{minipage}\hfill
\begin{minipage}{.25\textwidth} \centering
\begin{tikzpicture}[scale=0.47]
\begin{axis}[
    xlabel={Throughput [Ktrans/sec]},
    ylabel={Latency [ms]},
    xmin=0, xmax=13,
    ymin=0, ymax=500,
    xtick={3,6,9,12},
    ytick={100,200,300,400},
    legend pos=north west,
    ymajorgrids=true,
    grid style=dashed,
]

\addplot[
    color=red,
    mark=o,
    mark size=4pt,
    line width=0.5mm,
    ]
    coordinates {
    (0.304,8)(2.214,22)(4.413,36)(6.583,61)(8.783,94)(9.104,164)};
    
\addplot[
    color=green,
    mark=square,
    mark size=4pt,
    line width=0.5mm,
    ]
    coordinates {
    (0.304,6)(2.198,11)(4.374,18)(6.491,37)(8.711,58)(10.683,75)(10.934,144)};

\addplot[
    color=blue,
    mark=*,
    mark size=4pt,
    line width=0.5mm,
    ]
    coordinates {
   (0.074,2)(0.690,2)(0.823,7)(2.300,28)(3.711,46)(6.349,87)(9.830,173)(12.143,274)(12.411,452)
   };
   
\addplot[
    color=black,
    mark=triangle,
    mark size=4pt,
    line width=0.5mm,
    ]
    coordinates {
   (0.074,2)(0.690,19)(2.300,43)(4.111,78)(5.830,109)(7.113,152)(8.211,201)(8.511,403)
   };

\addlegendentry{APR-C}
\addlegendentry{F-Paxos}
\addlegendentry{SharPer}
\addlegendentry{AHL-C}

\end{axis}
\end{tikzpicture}

{\footnotesize (c) $80\%$ Cross-shard}
\end{minipage}\hfill
\begin{minipage}{.25\textwidth} \centering
\begin{tikzpicture}[scale=0.47]
\begin{axis}[
    xlabel={Throughput [Ktrans/sec]},
    ylabel={Latency [ms]},
    xmin=0, xmax=12,
    ymin=0, ymax=500,
    xtick={0,3,6,9},
    ytick={100,200,300,400},
    legend pos=north west,
    ymajorgrids=true,
    grid style=dashed,
]

\addplot[
    color=red,
    mark=o,
    mark size=4pt,
    line width=0.5mm,
    ]
    coordinates {
    (0.304,8)(2.214,22)(4.413,36)(6.583,61)(8.783,94)(9.104,164)};
    
\addplot[
    color=green,
    mark=square,
    mark size=4pt,
    line width=0.5mm,
    ]
    coordinates {
    (0.304,6)(2.198,11)(4.374,18)(6.491,37)(8.711,58)(10.683,75)(10.934,144)};

\addplot[
    color=blue,
    mark=*,
    mark size=4pt,
    line width=0.5mm,
    ]
    coordinates {
  (0.067,2)(0.590,4)(0.792,5)(2.023,34)(3.409,57)(5.450,101)(7.372,152)(9.452,210)(10.723,290)(10.932,470)};
   
\addplot[
    color=black,
    mark=triangle,
    mark size=4pt,
    line width=0.5mm,
    ]
    coordinates {
   (0.074,3)(0.582,14)(1.704,37)(3.281,82)(4.761,130)(5.404,154)(6.507,202)(7.373,281)(7.501,453)};

\addlegendentry{APR-C}
\addlegendentry{F-Paxos}
\addlegendentry{SharPer}
\addlegendentry{AHL-C}

\end{axis}
\end{tikzpicture}
{\footnotesize (d) $100\%$ cross-shard}
\end{minipage}
\caption{Increasing the Percentage of Cross-Shard Transactions in Networks with Crash-Only Nodes}
  \label{fig:cross-C}
\end{figure*}
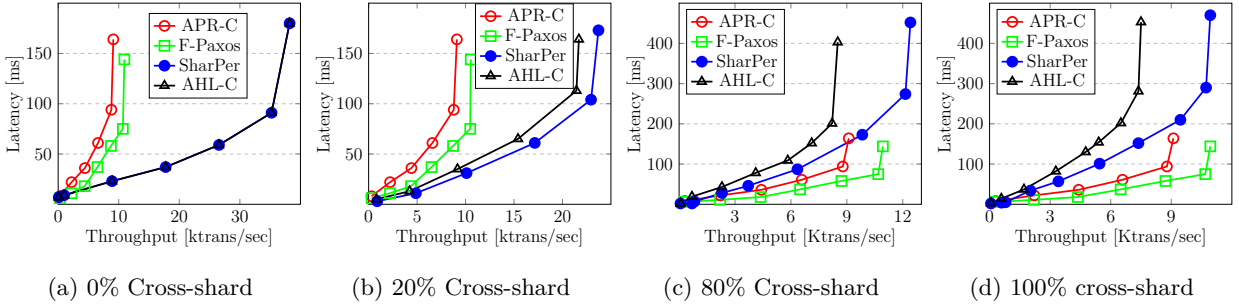

\subsection{Impact of Cross-Shard Transactions on Networks Consisting of Crash-Only Nodes}

In the first set of experiments, we measure the performance of \chain for
workloads with different percentages of cross-shard transactions where nodes are crash-only.
We compare \chain with the two main approaches for exploiting the availability of extra resources:
the active/passive replication technique and Fast Paxos \cite{lamport2006fast}.
We implemented two permissioned blockchain systems referred to as {\em APR-C} and {\em FPaxos}
where their consensus protocols follow the active/passive replication and Fast Paxos designs respectively.
In addition to \chain and these two systems, we also implemented a modified version of the
state of the art sharded permissioned blockchain system AHL \cite{dang2018towards}.
AHL has two novel aspects:
first, its intra-shard consensus protocol that uses trusted hardware to restrict the malicious behavior of nodes, and
second, its cross-shard consensus protocol where a reference committee uses 2PC to order the transactions.
Since the emphasis of the experiments is on cross-shard transactions, we implemented a modified version of AHL, called AHL-C
where the intra-shard transactions are processed similar to \chain,
however, the cross-shard transactions are performed similar to AHL \cite{dang2018towards}.
In this set of experiments,
since the nodes are crash-only, the reference committee uses Paxos \cite{lamport2001paxos}
to establish consensus.
Note that, since intra-shard consensus is pluggable, the trusted hardware technique can be employed in \chain as well.

We consider a network with $12$ nodes.
In \chain and AHL-C, the nodes are divided into four clusters where
each cluster consists of $3$ nodes and
uses Paxos with $f{=}1$ to establish consensus.
In AHL-C, a reference committee of three crash-only nodes is also considered to
establish consensus on the order of cross-shard transactions.
The data is also equally sharded into four shards.
In the APR-C blockchain system, $3$ nodes are used as the active replicas and
the execution results are sent to the remaining $9$ nodes whereas
FPaxos uses $4$ nodes ($3f+1$) to establish consensus and the results are sent to the remaining $8$ nodes.

We consider four different workloads with
(1) no cross-shard transactions,
(2) $20\%$ cross-shard transactions,
(3) $80\%$ cross-shard transactions, and
(4) $100\%$ cross-shard transactions.
We also assume that two (randomly chosen) shards are involved in each cross-shard transaction.
Note that since APR-C and FPaxos do not use sharding,
the percentage of cross-shard transactions does not affect their performance.
The load is also equally distributed among all the nodes.

As can be seen in Figure~\ref{fig:cross-C}(a), when there are no cross-shard transactions,
\chain is able to process $35230$ transactions with $91$ ms latency
before the end-to-end throughput is saturated (the penultimate point).
Note that in this setting, 
since there are no cross-shard transactions, each cluster orders and executes its transactions independently, thus
the throughput of the entire system will increase linearly with the increasing number of clusters.
Since for intra-shard transactions, AHL-C uses the same technique as \chain, its results are identical to \chain.
APR-C and FPaxos are also able to process $8800$ and $10700$ transactions with $95$ ms and $75$ ms latency respectively
before the end-to-end throughput is saturated (the penultimate points).
Note that since FPaxos establishes consensus in less number of phases, it has better performance than APR-C.
However, they both have much lower throughput in comparison to \chain
($25\%$ and $33\%$ of \chain at $60$ ms latency).
The results mainly demonstrate the effectiveness of employing the sharding technique in blockchain.

By increasing the percentage of cross-shard transactions to $20\%$ (Figure~\ref{fig:cross-C}(b)),
the throughput is reduced due to the overhead of
cross-shard transactions. In this setting, \chain is still able to process $23000$ transaction with $100$ ms latency
(the penultimate point)
whereas AHL-C processes $21000$ transactions at the same latency.
This is expected because
first, \chain, in contrast to AHL-C, is able to process non-overlapping cross-shard transactions in parallel, and
second, the cross-shard protocol of \chain involves less number of communication phases.
As mentioned before, since the sharding technique is not utilized by APR-C and FPaxos,
the percentage of cross-shard transactions does not affect their performance.

Similarly, increasing the percentage of cross-shard transactions to $80\%$ (Figure~\ref{fig:cross-C}(c))
and finally, $100\%$ (Figure~\ref{fig:cross-C}(d)) reduces
the peak throughput of \chain to $12300$ and $10500$, respectively.
Note that by increasing the percentage of cross-shard transactions, \chain still shows much better performance compare to AHL-C
($44\%$ better in their peak throughput with $100\%$ cross-shard transactions)
because \chain is still able to process non-overlapping cross-transactions in parallel and also needs less number of
communication phases.
In these two scenarios, since APR-C and FPaxos order the transactions using only three ($2f{+}1$) and four ($3f{+}1$) nodes,
their latency is lower than \chain.
Specially FPaxos processes transactions with significantly lower latency
due to its fast consensus routine.
However, since a large percentage of transactions is cross-shard,
\chain needs the participation of all involved clusters to order transactions and using sharding
has no significant advantage.

\subsection{Impact of Cross-Shard Transactions on Networks Consisting of Byzantine Nodes}

In the second set of experiments, we repeat the previous scenarios on networks with Byzantine nodes.
Similar to the previous section, we implement four permissioned blockchain systems:
(1) \chain,
(2) APR-B where its consensus protocol follows the active/passive replication technique
on Byzantine nodes),
(3) FaB where its consensus protocol follows Fast Byzantine Consensus protocol \cite{martin2006fast}
and uses $5f+1$ nodes (instead of $3f+1$) to establish consensus in two phases (instead of three as in PBFT), and
(4) AHL-B where its intra-shard transactions are processed using PBFT (similar to \chain) and
its cross-shard transactions follow AHL \cite{dang2018towards}.

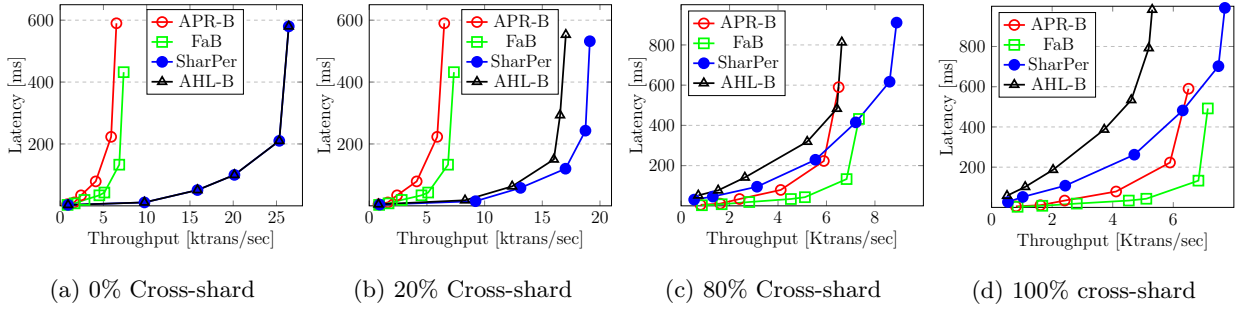
\begin{figure*}[t!]
\Large
\begin{minipage}{.25\textwidth}\centering
\begin{tikzpicture}[scale=0.47]
\begin{axis}[
    xlabel={Throughput [ktrans/sec]},
    ylabel={Latency [ms]},
    xmin=0, xmax=28,
    ymin=0, ymax=650,
    xtick={0,5,10,15,20,25},
    ytick={200,400,600},
    legend style={at={(axis cs:10,360)},anchor=south west}, 
    ymajorgrids=true,
    grid style=dashed,
]

\addplot[
    color=red,
    mark=o,
    mark size=4pt,
    line width=0.5mm,
    ]
    coordinates {
    (0.804,5)(1.614,11)(2.413,34)(4.112,79)(5.891,223)(6.503,590)};
    
\addplot[
    color=green,
    mark=square,
    mark size=4pt,
    line width=0.5mm,
    ]
    coordinates {
    (0.859,3)(1.675,8)(2.813,19)(4.531,34)(5.11,43)(6.83,133)(7.34,432)};

\addplot[
    color=blue,
    mark=*,
    mark size=4pt,
    line width=0.5mm,
    ]
    coordinates {
    (0.913,3)(9.742,11)(15.87,51)(20.14,100)(25.321,210)(26.417,580)};

\addplot[
    color=black,
    mark=triangle,
    mark size=4pt,
    line width=0.5mm,
    ]
    coordinates {
    (0.913,3)(9.742,11)(15.87,51)(20.14,100)(25.321,210)(26.417,580)};

\addlegendentry{APR-B}
\addlegendentry{FaB}
\addlegendentry{SharPer}
\addlegendentry{AHL-B}

\end{axis}
\end{tikzpicture}
{\footnotesize (a) $0\%$ Cross-shard}
\end{minipage}\hfill
\begin{minipage}{.25\textwidth} \centering
\begin{tikzpicture}[scale=0.47]
\begin{axis}[
    xlabel={Throughput [ktrans/sec]},
    ylabel={Latency [ms]},
    xmin=0, xmax=21,
    ymin=0, ymax=650,
    xtick={0,5,10,15,20},
    ytick={200,400,600},
    legend style={at={(axis cs:8,360)},anchor=south west}, 
    ymajorgrids=true,
    grid style=dashed,
]

\addplot[
    color=red,
    mark=o,
    mark size=4pt,
    line width=0.5mm,
    ]
    coordinates {
    (0.804,5)(1.614,11)(2.413,34)(4.112,79)(5.891,223)(6.503,590)};
    
\addplot[
    color=green,
    mark=square,
    mark size=4pt,
    line width=0.5mm,
    ]
    coordinates {
    (0.859,3)(1.675,8)(2.813,19)(4.531,34)(5.11,43)(6.83,133)(7.34,432)};

\addplot[
    color=blue,
    mark=*,
    mark size=4pt,
    line width=0.5mm,
    ]
    coordinates {
    (0.901,4)(9.213,15)(13.11,58)(17.013,120)(18.742,243)(19.120,532)};
  
\addplot[
    color=black,
    mark=triangle,
    mark size=4pt,
    line width=0.5mm,
    ]
    coordinates {
    (0.801,5)(8.304,18)(12.39,63)(16.003,150)(16.542,293)(17.04,553)};

\addlegendentry{APR-B}
\addlegendentry{FaB}
\addlegendentry{SharPer}
\addlegendentry{AHL-B}
 
\end{axis}
\end{tikzpicture}
{\footnotesize (b) $20\%$ Cross-shard}
\end{minipage}\hfill
\begin{minipage}{.25\textwidth} \centering
\begin{tikzpicture}[scale=0.47]
\begin{axis}[
    xlabel={Throughput [Ktrans/sec]},
    ylabel={Latency [ms]},
    xmin=0, xmax=10,
    ymin=0, ymax=1000,
    xtick={0,2,4,6,8},
    ytick={200,400,600,800},
    legend pos=north west,
    ymajorgrids=true,
    grid style=dashed,
]

\addplot[
    color=red,
    mark=o,
    mark size=4pt,
    line width=0.5mm,
    ]
    coordinates {
    (0.804,5)(1.614,11)(2.413,34)(4.112,79)(5.891,223)(6.503,590)};
    
\addplot[
    color=green,
    mark=square,
    mark size=4pt,
    line width=0.5mm,
    ]
    coordinates {
    (0.859,3)(1.675,8)(2.813,19)(4.531,34)(5.11,43)(6.83,133)(7.34,432)};

\addplot[
    color=blue,
    mark=*,
    mark size=4pt,
    line width=0.5mm,
    ]
    coordinates {
    (0.54,31)(1.31,46)(3.133,95)(5.543,229)(7.209,415)(8.605,617)(8.894,911)};

\addplot[
    color=black,
    mark=triangle,
    mark size=4pt,
    line width=0.5mm,
    ]
    coordinates {
    (0.71,52)(1.53,77)(2.631,141)(5.204,319)(6.431,483)(6.632,813)};

\addlegendentry{APR-B}
\addlegendentry{FaB}
\addlegendentry{SharPer}
\addlegendentry{AHL-B}

\end{axis}
\end{tikzpicture}
{\footnotesize (c) $80\%$ Cross-shard}
\end{minipage}\hfill
\begin{minipage}{.25\textwidth} \centering
\begin{tikzpicture}[scale=0.47]
\begin{axis}[
    xlabel={Throughput [Ktrans/sec]},
    ylabel={Latency [ms]},
    xmin=0, xmax=8,
    ymin=0, ymax=1000,
    xtick={0,2,4,6},
    ytick={200,400,600,800},
    legend pos=north west,
    ymajorgrids=true,
    grid style=dashed,
]

\addplot[
    color=red,
    mark=o,
    mark size=4pt,
    line width=0.5mm,
    ]
    coordinates {
    (0.804,5)(1.614,11)(2.413,34)(4.112,79)(5.891,223)(6.503,590)};
    
\addplot[
    color=green,
    mark=square,
    mark size=4pt,
    line width=0.5mm,
    ]
    coordinates {
    (0.859,3)(1.675,8)(2.813,19)(4.531,34)(5.11,43)(6.83,133)(7.14,492)};

\addplot[
    color=blue,
    mark=*,
    mark size=4pt,
    line width=0.5mm,
    ]
    coordinates {
    (0.54,27)(1.03,52)(2.433,108)(4.713,262)(6.312,482)(7.491,702)(7.703,992)};
   
\addplot[
    color=black,
    mark=triangle,
    mark size=4pt,
    line width=0.5mm,
    ]
    coordinates {
    (0.51,58)(1.11,102)(2.031,187)(3.720,387)(4.613,534)(5.191,792)(5.303,983)};

\addlegendentry{APR-B}
\addlegendentry{FaB}
\addlegendentry{SharPer}
\addlegendentry{AHL-B}

\end{axis}
\end{tikzpicture}
{\footnotesize (d) $100\%$ cross-shard}
\end{minipage}
\caption{Increasing the Percentage of Cross-Shard Transactions in Networks with Byzantine Nodes}
  \label{fig:cross-B}
\end{figure*}

We consider a network with $16$ nodes.
In \chain and AHL-B, the nodes are partitioned into $4$ clusters where
each cluster consists of four nodes and
uses PBFT protocol with $f=1$ to establish consensus on its transactions.
In addition to these 16 nodes, in AHL-B, a reference committee of four Byzantine nodes is also considered to
establish consensus on the order of cross-shard transactions.
In the APR-B blockchain system, $4$ nodes are used as the active replicas and
finally, FaB uses $6$ nodes ($5f+1$) to establish consensus.
Similar to the previous case, since APR-B and FaB do not use sharding,
the percentage of cross-shard transactions does not affect their performance.

As shown in Figure~\ref{fig:cross-B}(a), with no cross-shard transactions,
\chain is able to process more than $25000$ transactions with $200$ ms latency.
As before, since for intra-shard transactions, AHL-B uses the same technique as \chain,
the results of \chain and AHL-B are the same.
APR-B and FaB also process $5900$ and $6800$ transactions ($23\%$ and $27\%$ of \chain)
with $220$ ms and $130$ ms latency respectively.
Note that since transactions are processed in two phases (instead of three), FaB has lower latency in comparison to APR-B.

Increasing the percentage of cross-shard transactions to $20\%$, reduces the peak throughput of \chain to
$18700$ (with $240$ ms latency).
In this scenario and in comparison to AHL-B, \chain is able to
process $15\%$ more transactions (at their respective peak throughput) because of
the parallel ordering of cross-shard transactions and
establishing cross-shard consensus in less number of phases.
As mentioned before, since the sharding technique is not utilized by APR-B and FaB,
the percentage of cross-shard transactions does not affect their performance.
Note that with $20\%$ cross-shard transactions, the peak throughput of \chain is $320\%$ and $270\%$
of the peak throughput of APR-B and FaB respectively.

With $80\%$ cross-shard transactions, the peak throughput of \chain reduces to $8600$
which is still $34\%$ higher than the peak throughput of AHL-B ($6400$) due to parallel processing of 
non-overlapping cross-shard transactions.
Finally, when all transactions are cross-shard,
\chain is able to process $7500$ transactions with $700$ ms latency whereas
AHL-B processes $5000$ transactions ($67\%$ of \chain) with the same latency.
In the last two scenarios ($80\%$ and $100\%$ cross-shard transactions), 
because of the high percentage of cross-shard transactions, using sharding techniques
has no significant advantage and
since APR-B and FaB rely on only four ($3f{+}1$) and six ($5f{+}1$) nodes to order transactions respectively,
their latency is lower than \chain.
However, in \chain, simultaneous processing of non-overlapping transactions results in 
improved throughput.

\subsection{Performance with Different Number of Nodes}

In the last set of experiments, we measure the performance of \chain in networks
with different number of nodes.
We measure the performance of \chain in a network including $6$, $9$, $12$, and $15$ crash-only nodes as well as
$8$, $12$, $16$ and $20$ Byzantine nodes ($2$, $3$, $4$ and $5$ clusters).
The workloads also include
$90\%$ intra- and $10\%$ cross-shard transactions
(the typical settings in partitioned databases \cite{thomson2012calvin} \cite{taft2014store}).

\begin{figure}[t]
\centering
\large
\begin{minipage}{.23\textwidth}\centering
\begin{tikzpicture}[scale=0.45]
\begin{axis}[
    xlabel={Throughput [ktrans/sec]},
    ylabel={Latency [ms]},
    xmin=0, xmax=38,
    ymin=0, ymax=230,
    xtick={0,6,12,18,24,30,36},
    ytick={50,100,150,200},
    legend columns=3,
    legend style={at={(axis cs:-4,250)},anchor=south west}, 
    ymajorgrids=true,
    grid style=dashed,
]

\addplot[
    color=red,
    mark=o,
    mark size=4pt,
    line width=0.5mm,
    ]
    coordinates {
    (0.804,5)(1.614,11)(2.413,34)(4.112,79)};
    
\addplot[
    color=green,
    mark=square,
    mark size=4pt,
    line width=0.5mm,
    ]
    coordinates {
    (0.859,3)(1.675,8)(2.813,19)(4.531,34)(5.11,43)(6.83,133)};

 \addplot[
    color=black,
    mark=+,
    mark size=4pt,
    line width=0.5mm,
    ]
    coordinates {
    (0.861,3)(2.813,9)(6.014,24)(9.811,53)(11.059,75)(12.883,119)};

\addplot[
    color=magenta,
    mark= square*,
    mark size=4pt,
    line width=0.5mm,
    ]
    coordinates {
    (0.890,4)(4.013,11)(9.452,22)(13.911,49)(17.302,79)(20.311,105)(20.619,171)};

\addplot[
    color=blue,
    mark=*,
    mark size=4pt,
    line width=0.5mm,
    ]
    coordinates {

  (0.888,3)(5.113,10)(11.145,27)(19.319,56)(23.802,71)(28.152,106)(29.819,164)};

\addplot[
    color=violet,
    mark=triangle,
    mark size=4pt,
    line width=0.5mm,
    ]
    coordinates {
  (0.911,2)(6.413,13)(14.045,23)(24.019,41)(29.302,58)(36.852,102)(37.219,184)};

\addlegendentry{APR-C}
\addlegendentry{FPaxos}
\addlegendentry{$6$ nodes}
\addlegendentry{$9$ nodes}
\addlegendentry{$12$ nodes}
\addlegendentry{$15$ nodes}
 
\end{axis}
\end{tikzpicture}
{\footnotesize (a) Crash-Only Nodes}
\end{minipage}
\begin{minipage}{.23\textwidth} \centering
\begin{tikzpicture}[scale=0.45]
\begin{axis}[
    xlabel={Throughput [ktrans/sec]},
    ylabel={Latency [ms]},
    xmin=0, xmax=30,
    ymin=0, ymax=650,
    xtick={0,5,10,15,20,25},
    ytick={200,400,600},
   legend columns=3,
   legend style={at={(axis cs:-3,700)},anchor=south west}, 
    ymajorgrids=true,
    grid style=dashed,
]

\addplot[
    color=red,
    mark=o,
    mark size=4pt,
    line width=0.5mm,
    ]
    coordinates {
    (0.804,5)(1.614,11)(2.413,34)(4.112,79)(5.891,223)(6.503,590)};
    
\addplot[
    color=green,
    mark=square,
    mark size=4pt,
    line width=0.5mm,
    ]
    coordinates {
    (0.859,3)(1.675,8)(2.813,19)(4.531,34)(5.11,43)(6.83,133)(7.34,432)};

 \addplot[
    color=black,
    mark=+,
    mark size=4pt,
    line width=0.5mm,
    ]
    coordinates {
    (0.867,3)(4.633,16)(7.002,49)(8.930,98)(10.913,224)(11.113,363)};

\addplot[
    color=magenta,
    mark=square*,
    mark size=4pt,
    line width=0.5mm,
    ]
    coordinates {
    (0.913,4)(7.212,16)(10.809,55)(13.554,105)(16.218,189)(16.710,470)};

\addplot[
    color=blue,
    mark=*,
    mark size=4pt,
    line width=0.5mm,
    ]
    coordinates {
    (0.893,3)(9.533,14)(14.402,53)(18.39,113)(22.043,220)(22.340,603)};

\addplot[
    color=violet,
    mark=triangle,
    mark size=4pt,
    line width=0.5mm,
    ]
    coordinates {
    (0.917,3)(13.002,13)(17.423,52)(22.011,108)(27.408,241)(27.903,480)};

\addlegendentry{APR-B}
\addlegendentry{FaB}
\addlegendentry{$8$ nodes}
\addlegendentry{$12$ nodes}
\addlegendentry{$16$ nodes}
\addlegendentry{$20$ nodes}

\end{axis}
\end{tikzpicture}
{\footnotesize (b) Byzantine Nodes}
\end{minipage}
\caption{Increasing the Number of Nodes}
  \label{fig:node}
\end{figure}
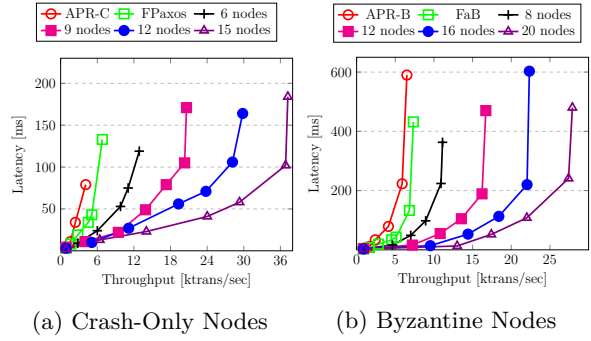

As can be seen in Figure~\ref{fig:node}(a),
when nodes follow the crash failure model,
by increasing the number of nodes (clusters)
the throughput of the system increases almost linearly.
This is expected because $90\%$ of transactions are intra-shard transactions and, as shown earlier,
for intra-shard transactions,
the throughput of the entire system will increase linearly with the increasing number of clusters.
In addition, since cross-shard transactions access two clusters, by increasing the number of clusters,
the chance of parallel processing of such transactions increases.
As shown in Figure~\ref{fig:node}(a), in the settings with five clusters,
\chain is able to process $37000$ transactions with $100$ ms latency.
Since increasing the number of nodes does not significantly affect the performance of
APR-C and FPaxos systems, their performance will be similar to what is reported
in Figure~\ref{fig:cross-C}. However, as can be seen, in a network consisting of $6$ nodes ($50\%$ more nodes than FPaxos)
\chain processes upto $11060$ transactions ($88\%$ more than FPaxos) with the same ($75$ ms) latency.

Similarly, when nodes follow the Byzantine failure model,
increasing the number of clusters enhances the overall throughput of \chain, as shown in Figure~\ref{fig:node}(b).
In this scenario, \chain can process more than $27000$ transactions with $240$ ms latency on a network with five clusters.
Furthermore, in a network with $8$ nodes and with $200$ ms latency,
\chain, using only $33\%$ more nodes, is able to process $58\%$ more transactions than FaB.
This set of experiments demonstrates the scalability of \chain as the number of clusters increases.

\subsection{Discussion}

Overall, the evaluation results can be summarized as follow.

First, in typical settings where workloads include low percentage (less than $20\%$) of cross-shard transactions,
\chain demonstrates better performance with both crash-only and Byzantine nodes in comparison to other approaches.
The performance of \chain is better than AHL \cite{dang2018towards} because of the cross-shard consensus routine of
\chain that, in contrast to AHL, can order cross-shard transactions with non-overlapping clusters in parallel.
The performance of \chain is much (three to four times) better than both FPaxos (FaB) and active/passive replication (APR-C and APR-B)
since \chain uses the sharding technique and is able to process intra-shard transactions of different clusters in parallel
whereas in both FPaxos (FaB) and active/passive replication, transactions are processed sequentially.
Furthermore, and as shown in Figure~\ref{fig:node}, the performance of \chain enhances semi-linearly
with the increasing number of clusters, which clearly demonstrates the scalability of \chain.

Second, in settings with high percentage of cross-shard transactions, using sharding techniques
has no significant advantage.
As a result, in the presence of extra nodes, using FPaxos (FaB) and active-passive replication (APR-C and APR-B)
results in better performance (specially less latency).
Note that as mentioned before, the typical settings in partitioned database systems
includes only $10\%$ cross-shard transactions \cite{thomson2012calvin} \cite{taft2014store}.

\section{Related Work}\label{sec:related}

State machine replication (SMR) is a technique for implementing a fault-tolerant service
by replicating servers \cite{lamport1978time}.
Paxos \cite{lamport2001paxos} generalizes SMR to support crash failures and
guarantees safety in an asynchronous network using $2f{+}1$ nodes
despite the simultaneous crash failure of any $f$ nodes.
If the number of available nodes is more than $2f+1$,
Lamport \cite{lamport2006fast} and  Brasileiro et al. \cite{brasileiro2001consensus} 
can utilize $f$ more nodes to reduce one phase of communication.
Alternatively, the extra nodes can become {\em passive} replicas and only be informed
about the execution results, so that their copies of the ledger become up to date.

Byzantine fault tolerance refers to servers that behave arbitrarily
after the seminal work by Lamport, et al. \cite{lamport1982byzantine}.
Practical Byzantine fault tolerance protocol (PBFT) \cite{castro1999practical}
is the most well-known state machine replication protocol that
guarantees safety in an asynchronous network using $3f+1$ nodes from which
$f$ might be malicious.
Consensus protocols explore a spectrum of performance trade-offs between the number of required participants,
number of phases/messages (latency), and message complexity.
In the presence of more than $3f+1$ nodes, similar to crash fault-tolerant protocols,
one solution is to use the active/passive replication technique where
only $3f+1$ {\em active} replicas establish consensus on the order of requests, execute
the requests, and send the execution results to the passive replicas.
FaB \cite{martin2006fast}, Bosco \cite{song2008bosco}, and Zyzzyva5 \cite{kotla2007zyzzyva},
on the other hand,
use additional replicas to reduce the delay of request processing, e.g.,
FaB \cite{martin2006fast} uses $5f+1$ replicas to establish consensus
on the order of requests in two phases instead of three as in PBFT.
While both crash fault-tolerant protocol Fast Paxos \cite{lamport2006fast} and
Byzantine fault-tolerant protocol FaB \cite{martin2006fast}
use extra nodes ($f$ and $2f$ respectively) to reduce the latency of the system,
if the number of extra nodes is more than that (the typical case in Blockchain systems),
such nodes cannot be utilized and in the best case scenario the extra nodes become passive replicas.
In \chain, however, the nodes are partitioned into clusters of $2f+1$ crash-only or $3f+1$ Byzantine nodes
to process transactions in parallel.
To reduce the number of replicas, some approaches rely on a trusted component
\cite{chun2007attested, veronese2013efficient, correia2004tolerate, distler2016resource, kapitza2012cheapbft}
that prevents a faulty node from sending conflicting
messages to different nodes without being detected.
SBFT \cite{gueta2018sbft} and Hotstuff \cite{yin2019hotstuff} attain
linear communication overhead by increasing the number of communication phases
and using advanced encryption techniques, e.g., signature aggregation \cite{boneh2004short}.
Finally, MultiBFT \cite{gupta2019scaling} uses multiple parallel primary nodes to parallelize transaction processing and
hence improve performance.

Replication techniques, both eager, i.e., all replicas are synchronized as part of atomic transactions, and
lazy, i.e., updates are propagated asynchronously to other nodes
after committing transactions \cite{holliday1999database} \cite{breitbart1999update},
have been extensively used by distributed databases to enhance reliability and availability in
networks consisting of crash-only nodes \cite{gray1996dangers}.

A permissioned blockchain consists of a set of known, identified nodes that might not fully trust each other.
In permissioned blockchains,
traditional consensus protocols can be used to order the requests \cite{cachin2016architecture}.
Existing permissioned blockchains,
e.g., Tendermint \cite{kwon2014tendermint}, Quorum \cite{morgan2016quorum}, Fabric \cite{androulaki2018hyperledger}, Parblockchain \cite{amiri2019parblockchain}, Fast Fabric \cite{gorenflo2019fastfabric}, ResilientDB \cite{gupta2020resilientdb}, and Caper \cite{amiri2019caper},
differ mainly in their ordering routines.
Quorum \cite{morgan2016quorum}, is an Ethereum-based \cite{ethereum17} permissioned blockchain,
that uses a Raft-like \cite{ongaro2014search} protocol to order transactions.
Single-channel Hyperledger Fabric \cite{androulaki2018hyperledger} deploys different applications on the same channel and
leverages parallelism by executing the transactions of those applications simultaneously.
In Fabric, fault-tolerant protocols are pluggable. In addition, Fabric supports non-deterministic execution.
However, since the transactions of a block are executed in parallel and then ordered and validated,
Fabric performs poorly on workloads with high-contention, i.e., many {\em conflicting transactions} in a block.
To support conflicting transactions, in Parblockchain \cite{amiri2019parblockchain},
a dependency graph is generated in the ordering phase and 
transactions are executed in parallel in the execution phase following the generated dependency graph.
In \chain, however, since each block includes a single transaction, transactions will not conflict with each other.
Caper \cite{amiri2019caper} is another permissioned blockchain that is introduced to support collaborative applications.
In Caper, transactions are either internal, which are maintained by a single application,
or cross-application, which are maintained by all applications.
Each application also maintains two types of private and public data.
\chain, in contrast to Caper, is able to handle transactions that access a subset of clusters (i.e. applications).
In addition, in \chain, both intra- and cross-shard transactions access the same data.

Scalability is the ability of a system to process an increasing number of transactions
by adding resources to the system.
While the Visa payment service is able to handle on average $2000$ transactions per second,
Bitcoin and Ethereum can handle at most $7$ and $15$ transactions per second respectively.
To address the scalability issue different techniques have been proposed.
Off-chain (layer two) solutions, which are built on top of the main-chain, do not increase the throughput of the protocol,
rather move a portion of the transactions off the chain.
For example, in Lightning Networks \cite{miller2019sprites}\cite{poon2016bitcoin},
assets are transferred between two different clients via a network of micro-payment channels instead of
the main blockchain.
While off-chain solutions increase the throughput of the system,
they suffer from security issues, e.g., denial-of-service attacks.

In general security, decentralization, and performance are known as the {\em scalability trilemma} in blockchain systems.
Security requires resistance to threats such as the denial-of-service attacks, $51\%$ attack, or Sybil attacks;
decentralization means no single entity can hijack the chain, censor it, or introduce changes in governance; and
performance is the ability to handle thousands of transactions per second.

On-chain (layer one) solutions, on the other hand, increase the throughput of the main chain.
Layer one solutions are categorized into vertical and horizontal techniques.
In vertical scalability, more power is added to each node to perform more tasks.
One trivial solution is to increase the block size which results in processing more transactions at once,
thus enhancing performance.
Increasing the block size, however, increases both the propagation time and the verification time of the block which
makes operating full nodes more expensive, and this in turn could cause less decentralization in the network.

Horizontal techniques, on the other hand, increase the number of nodes in the network to process more transactions.
However, most blockchain systems require every transaction to be processed by every single node in the network.
As a result, increasing the number of nodes, does not necessarily enhance the performance of the system.

Another horizontal solution to enhance the scalability of blockchain systems is sharding.
Partitioning the data into multiple shards that are maintained by different subsets of nodes
is a proven approach to enhance the scalability of databases \cite{corbett2013spanner}.
Data sharding techniques are commonly used in globally distributed databases such as
H-store \cite{kallman2008h}, Calvin \cite{thomson2012calvin},
Spanner \cite{corbett2013spanner}, Scatter \cite{glendenning2011scalable}, Google's Megastore \cite{baker2011megastore},
Amazon's Dynamo \cite{decandia2007dynamo}, Facebook's Tao \cite{bronson2013tao}, and E-store \cite{taft2014store}.
In such systems servers (nodes) are assumed to be crash-only and a coordinator node is used to process crash-shard transactions.
Agrawal et al. \cite{agrawal2015taxonomy} categorize sharded, replicated database systems into
replicated object systems, e.g., Spanner \cite{corbett2013spanner} and replicated transaction systems,
e.g., replicated commit protocol \cite{mahmoud2013low}.
\chain is inspired by distributed database systems and has applied the sharding technique to the blockchain domain.
Furthermore, \chain proposes consensus protocol for network consisting of Byzantine nodes
and introduces a flattened cross-shard consensus protocol instead of a coordinator-based one.

Sharding techniques have been used in both permissionless, e.g.,
Elastico \cite{luu2016secure}, OmniLedger \cite{kokoris2018omniledger}, and
Rapidchain \cite{zamani2018rapidchain}, and
permissioned blockchain systems, e.g., multi-channel Fabric \cite{androulaki2018channels}, AHL \cite{dang2018towards},
Cosmos \cite{cosmos18}, and RSCoin \cite{george2015centrally} to improve scalability.
In Elastico \cite{luu2016secure}, nodes randomly join different committees by solving some PoW puzzle.
Committees, then, run PBFT \cite{castro1999practical}
individually to reach consensus on the order of intra-shard transactions.
Finally, a leader committee verifies the transactions
that are ordered by committees and creates a global block.
In Elastico, the blockchain ledger is maintained by all nodes and
cross-shard transactions are not supported.
In addition, while running PBFT among hundreds of nodes decreases the performance of the protocol,
reducing the number of nodes within each shard increases
the failure probability \cite{kokoris2018omniledger}.
The considerable overhead and latency in re-configuring committees, which is needed in every epoch, and
the possibility to bias the randomness, which might result in compromising the committee selection process by malicious nodes,
are some of the other drawbacks of Elastico \cite{zamani2018rapidchain}.

OmniLedger \cite{kokoris2018omniledger} attempts to fix some of the drawbacks of Elastico by
introducing a more secure method to assign nodes to committees and
proposing an atomic protocol for cross-shard transactions.
The intra-shard consensus protocol of OmniLedger uses a variant of ByzCoin \cite{kogias2016enhancing} and
assumes partially-synchronous channels to achieve fast consensus.
However, it relies on a client to participate actively
and coordinate a lock/unlock protocol to process cross-shard transactions which,
as shown in \cite{dang2018towards}, might result in blocking issues.
Furthermore, as mentioned in \cite{zamani2018rapidchain},
OmniLedger is vulnerable to denial-of-service (DoS) attacks.
In multi-channel Fabric \cite{androulaki2018hyperledger}\cite{androulaki2018channels},
as explained in Section~\ref{sec:intro},
processing cross-shard transactions, in contrast to \chain, requires
either the existence of a trusted channel among the participants or 
an atomic commit protocol (inspired by two-phase commit) \cite{androulaki2018channels}.
Similarly, in Cosmos \cite{cosmos18}, interacting chains in any Inter-Blockchain Communication
must be aware of the state of each other which requires
establishing a bidirectional trusted channel between two blockchains.

AHL \cite{dang2018towards} employs a trusted hardware 
(the technique that is presented in \cite{chun2007attested, veronese2013efficient, veronese2010ebawa})
to restrict the malicious behavior of nodes which results in committees of $2f+1$ nodes (instead of $3f+1$).
The system also relies on an extra set of nodes, called a reference committee, to process cross-shard transactions
using the classic two-phase commit (2PC) and two-phase locking (2PL) protocols where the reference committee
plays the coordinator role.
The system, however, suffers from several drawbacks.
First, running fault-tolerant protocols among $80$ nodes results in high latency.
Second, the protocol requires an extra set of nodes to form the reference committee resulting
in significant communication overhead between nodes and the reference committee.
Finally, since a single reference committee processes cross-shard transactions,
the protocol is not able to process cross-shard transactions with non-overlapping clusters in parallel.
In \chain and in contrast to AHL, there is no need for an extra set of nodes
to process cross-application transactions.
In addition, cross-shard transactions are ordered in only three communication phases.
Furthermore, cross-shard transactions with non-overlapping committees can be processed simultaneously.
Note that since intra-shard consensus is pluggable, 
the trusted hardware technique can be employed
to decrease the number of required nodes within each cluster.

Our work is also related to blockchain systems with directed acyclic graph structure.
The DAG structure is mainly used to increase the throughput of the system 
by exploiting the parallel construction of blocks resulting in the parallel execution of transactions in different blocks.
In such a structure, the blocks (transactions) that are independent of each other
can be appended to the ledger simultaneously.
Since in a DAG structure, the blocks are constructed in parallel,
existing permissionless blockchain systems
present different techniques to prevent (resolve) the double spending problem.
In Byteball \cite{churyumov2016byteball}, a set of privileged users, called witnesses,
determines a total order on the DAG to prevent double spending, whereas,
in Iota \cite{popov2016tangle}, the number of descendant transactions
is used to commit a transaction and abort the other one.
Vegvisir \cite{karlsson2018vegvisir}, which is designed for IoT environments,
Ghost \cite{sompolinsky2013accelerating},
Inclusive protocol \cite{lewenberg2015inclusive}, DagCoin \cite{lerner2015dagcoin}, 
Phantom \cite{sompolinsky2018phantom}, Spectre \cite{sompolinsky2016spectre},
MeshCash \cite{bentov2017tortoise}, and Hashgraph \cite{baird2016swirlds} are some of the other DAG structured
blockchain systems.
\chain is a permissioned blockchain system that establishes consensus on the order of transactions
using traditional fault-tolerant protocols Paxos and PBFT,
therefore, the forking or double spending problem never occurs.
In \chain and in contrast to all these blockchains, since intra-shard transactions of different clusters
access disjoint data shards, they can be processed simultaneously which results in lower latency and higher throughput.
\section{Conclusion}\label{sec:conc}

In this paper, we proposed \chain, a permissioned blockchain system
which is designed specifically for networks with a very high percentage of non-faulty nodes
($N \gg 3f+1$ for Byzantine or $N \gg 2f+1$ for crash-only nodes). 
\chain utilizes the extra resources by
partitioning the nodes into clusters of $3f+1$ Byzantine (or $2f+1$ crash-only) nodes and
processing the transactions on different clusters in parallel.
The blockchain ledger in \chain is formed as a directed acyclic graph which in not maintained by any node.
Nodes of each cluster indeed maintain a view of the blockchain ledger including
the intra-shard transactions of the cluster as well as
the cross-shard transactions that the cluster is involved in.
A flattened consensus protocol is also introduced to order cross-shard transactions without relying
on an extra set of nodes or trusted participants.
Furthermore, \chain is able to process cross-shard transactions with non-overlapping clusters in parallel.
Our experiments show that
in workloads with low percentage of cross-shard transactions (typical settings),
\chain demonstrates better performance with both crash-only and Byzantine nodes in comparison to other approaches and
the throughput of \chain will increase semi-linearly by increasing the number of clusters.
\balance

\bibliographystyle{abbrv}
\bibliography{main}

\end{document}